%% file: xenon100a.tex
\documentclass[a4paper,11pt]{article}
\usepackage{amsmath,amssymb,amsfonts}
\usepackage[numbers,sort&compress,merge]{natbib}
\usepackage{graphicx}
\usepackage{rotating}
\usepackage[hang,bf,nooneline]{subfigure}
\usepackage[hang,bf,nooneline]{caption}
\input{config}
\begin{document}
%\begin{frontmatter}
\begin{center}

%\Large \textbf{ Constraints from direct dark matter searches and LHC limits on SUSY  and Higgs particles  on Supersymmetry}

\Large \textbf{Where is SUSY? }
\vspace{10mm}

\large

C. Beskidt$^a$, W. de Boer$^{a}$\footnote{Email: wim.de.boer@kit.edu}, D.I. Kazakov$^{b,c}$, F. Ratnikov$^{a,c}$

\normalsize
\vspace{5mm}
$^a$ \textit{Institut f\"ur Experimentelle Kernphysik,
Karlsruhe Institute of Technology,\\ P.O. Box 6980, 76128 Karlsruhe, Germany}

\vspace{5mm}
$^b$ \textit{Bogoliubov Laboratory of Theoretical Physics, Joint Institute for Nuclear Research,\\
141980, 6 Joliot-Curie, Dubna, Moscow Region, Russia}

\vspace{5mm}
$^c$ \textit{Institute for Theoretical and Experimental Physics,\\
117218, 25 B.Cheremushkinskaya, Moscow, Russia}

\vspace{30mm} \textbf{Abstract} \vspace{5mm}

\begin{minipage}[c]{12cm}

\textit{
The direct searches for Superymmetry at colliders can be complemented by direct searches for dark matter (DM) in underground experiments, if one assumes the Lightest Supersymmetric Particle (LSP) provides the dark matter of the universe.
It will be shown that within the Constrained minimal Supersymmetric Model (CMSSM) the direct searches for DM  are complementary to direct LHC searches for SUSY and Higgs particles using analytical formulae.  A combined excluded region from LHC, WMAP and XENON100 will be provided, showing that within the CMSSM gluinos  below 1 TeV  and LSP masses below 160 GeV are excluded ($m_{1/2} > 400 GeV$) independent of the squark masses. 
}
\end{minipage}

\end{center}

\thispagestyle{empty}
\setcounter{page}{0}
%logic:

\section{Introduction}

Supersymmetry (SUSY) is the leading candidate for physics beyond the SM, since the Lightest Supersymmetric Particle (LSP) has all the properties expected for the Weakly Interacting Massive Particles (WIMPS)
of the dark matter \cite{Kolb:1990vq,Jungman:1995df,Bertone:2004pz}, which is known to make up more than 80\% of the matter in the universe \cite{Komatsu:2010fb}.
Unfortunately, no  supersymmetric particles have  been observed so far, even at the highest energy of the LHC \cite{Aad:2011qa,Chatrchyan:2011zy,Khachatryan:2011tk,Chatrchyan:2011bj,Collaboration:2011ida,Chatrchyan:2011qs,Chatrchyan:2011ah,Chatrchyan:2011wba,Chatrchyan:2011wc,Chatrchyan:2011ff,Aad:2011hh,Aad:2011xm,Aad:2011ks}, nor have WIMPS been observed beyond doubt in elastic scattering of WIMPS on nuclei, as pursued with direct searches in underground detectors, like CDMS,  EDELWEISS and XENON100, which give upper limits on the elastic scattering cross section typically between $10^{-43}$ and  $10^{-44}$ cm$^2$ \cite{Ahmed:2011gh,Aprile:2011hi}. The direct DM detection experiments, the search for SUSY and the search for Higgs particles all determine complementary excluded regions in the parameter space of supersymmetric models, like the Constrained Minimal Supersymmetric Model (CMSSM), see \cite{Chamseddine:1982jx,Kolda:1994ab} and reviews, e.g. \cite{Haber:1984rc,deBoer:1994dg,Martin:1997ns,Kazakov:2010qn}. In this model all the arguments in favour of Supersymmetry, like the unification of the coupling constants at the GUT (Grand Unified Theory) scale with SUSY masses in the TeV range \cite{Amaldi:1991cn} and electroweak symmetry breaking (EWSB) by radiative corrections \cite{Inoue:1982pi,Gladyshev:1996fx}, are implemented. In the CMSSM one furthermore assumes that  the masses of spin 0 (spin 1/2) particles are unified at the GUT scale with values $\mzero (\mhalf)$. So  the many parameters of SUSY models  are reduced to only 4: the two mass parameters $\mzero$, $\mhalf$ and two parameters related to the Higgs sector: the trilinear coupling at the GUT scale $A_0$, and \tb, the ratio of the vacuum expectation values of the two neutral components of the two Higgs doublets. Electroweak symmetry breaking (EWSB) fixes the scale of $\mu$, so only its sign is a free parameter.
  The positive sign is taken, as
suggested by the small deviation of the SM prediction from the muon anomalous moment, see e.g. \cite{deBoer:2001nu}. 
% Several heavy flavour physics observables  are enhanced by large values of $\tb$  as well (see e.g. \cite{Isidori:2007jw} and references therein). Especially, the FCNC decay of $B_s\rightarrow \mu\mu$, which proceeds in the SM via loops involving top quarks and W-bosons, is enhanced in SUSY by $\tan^6\beta$. Although the relic density constraint prefers large $\tb$ in most of the parameter region \cite{Beskidt:2010va}, the correlation with the trilinear coupling can be used to fulfill the experimental limits on $B_s\rightarrow \mu\mu$ \cite{Beskidt:2011qf}.

Within the CMSSM the direct searches for SUSY particles (''sparticles'') at the LHC, the direct DM searches and the relic density, as obtained from  cosmological observations, are related and one can combine them to see which region of the supersymmetric parameter space is excluded, if one includes all constraints. Such combinations have been pursued by many different groups  either using a frequentist approach by maximizing a likelihood  or using random sampling techniques of the parameter space, see e.g. \cite{Buchmueller:2011ki,Buchmueller:2011sw,Bertone:2011nj,Allanach:2011ut,
Allanach:2011wi,Farina:2011bh,Strumia:2011dy,Akula:2011dd,Trotta:2008bp,Akrami:2009hp,Feroz:2008wr,Sekmen:2011cz} and references therein. 

The sampling techniques are dependent on the prior, which leads to an additional, non-quantifiable uncertainty in the excluded or allowed regions, see e.g. \cite{Feroz:2011bj} for a recent discussion and references therein.  We believe this uncertainty is due to the high correlations between three of the four parameters, as we discussed in two previous papers: $\mzero$ is highly correlated with \tb~ because of the relic density constraint, which requires large \tb~ in most of the parameter space except for the narrow co-annihilation regions at low and large $\mzero$ \cite{Beskidt:2010va} and the trilinear coupling is highly correlated with \tb ~ because of heavy flavour constraints, notably the $\bsmm$ constraint \cite{Beskidt:2011qf}. Note that  $\bsmm$ is not only proportional to $\tan^6\beta$, so it tends to become above the present upper limit for large \tb, but it can be strongly reduced by the appropriate value of the trilinear coupling to values even below the SM value by negative interferences \cite{Beskidt:2011qf}.

Such strong correlations lead to
likelihood "spikes" in the parameter region, where three of the four parameters have to have specific correlated values. Although the likelihood of such narrow regions is high, they are either not found in methods based on stepping techniques or their probability is given a different weight because of its "low posterior mass".
To cope with the strong correlations we use a multistep fitting technique, defined by fitting the parameters with the strongest correlation first, i.e. we fit first $\tb$ an $A_0$ for each pair of the mass parameters  $\mzero$ and $\mhalf$ by  minimizing the $\chi^2$ with the program Minuit\cite{James:1975dr} or by using a Markov Chain Monte Carlo or multinest fitting technique to find the maximum likelihood, while integrating over the nuisance parameters.    Both the $\chi^2$ minimization and Markov Chain sampling of the parameter space give practical identical results in this multistep fitting technique and no prior dependence has been found, since for each point in the  ($\mzero$,$\mhalf$) grid there is  a unique solution for $\tb$, mainly from the relic density constraint, and $A_0$ can be fine-tuned to further minimize $\chi^2$. Furthermore the multistep fitting technique is  fast, since initially only two parameters are fitted for each  point in the  ($\mzero$,$\mhalf$) grid. Once the SUSY parameters have been fitted, one should also vary the SM parameters or marginalize over them, like the top and bottom mass and the strong coupling constant. However, these are highly correlated with the SUSY parameters, so if one   repeats the fit with  different values of the SM parameters or marginalizes over them one usually finds the same fit probability with slightly different values of the SUSY parameters. The SM parameters are  given in the Particle Data Book \cite{Nakamura:2010zzi}; we use $m_{top}^{pole}=172.5\pm 1.3$ and $m_b(m_b)^{\overline{MS}}=4.25\pm 0.2$ GeV for the heavy quark  masses  and  $\alpha_s=0.1172\pm 0.02$ for the strong coupling constant.

 All observables discussed below
were calculated with the public code micrOMEGAs 2.4.1
\cite{Belanger:2010pz,Pukhov:2010px} combined with Suspect 2.41 as  mass
spectrum calculator \cite{Djouadi:2002ze}.

The purpose of  this letter is twofold: we show that the data from the LHC, WMAP and XENON100  lead to complementary excluded regions and  discuss the reasons why by giving analytical approximations. We furthermore combine the newest data and show that a lower limit of $\mhalf$ of 400 GeV can be obtained independent of $\mzero$, which implies a lower limit of 160 GeV on the LSP and 1 TeV on the gluino.   
Our results differ from similar analysis referred to above and we discuss possible reasons.
We start by discussing the observations and the excluded regions of each observation separately.
\begin{figure}
 \begin{center}
 \includegraphics[width=0.24\textwidth,height=0.1\textwidth]{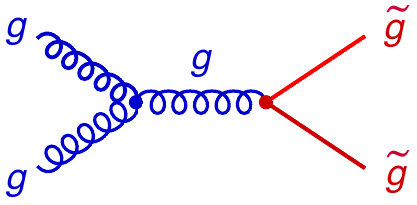}
\includegraphics[width=0.24\textwidth,height=0.1\textwidth]{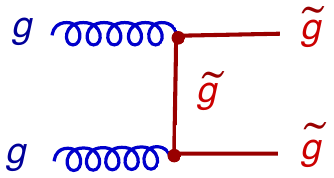}
\includegraphics[width=0.24\textwidth,height=0.1\textwidth]{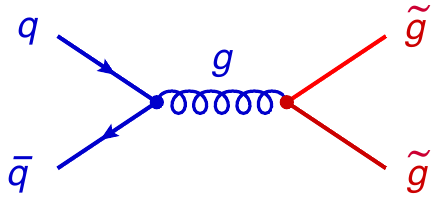}
\includegraphics[width=0.24\textwidth,height=0.1\textwidth]{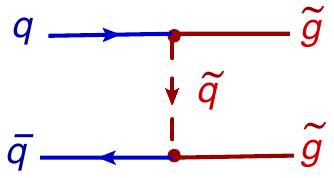}\\[4mm]
\includegraphics[width=0.24\textwidth,height=0.1\textwidth]{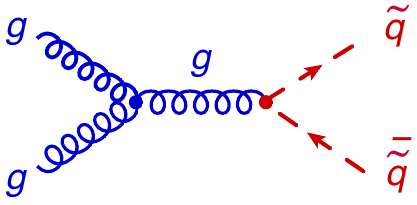}
\includegraphics[width=0.24\textwidth,height=0.1\textwidth]{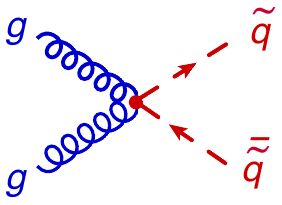}
\includegraphics[width=0.24\textwidth,height=0.1\textwidth]{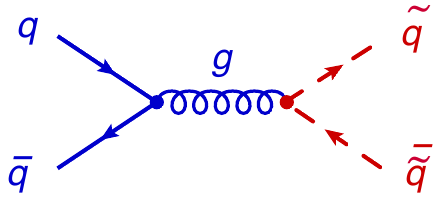}
\includegraphics[width=0.24\textwidth,height=0.1\textwidth]{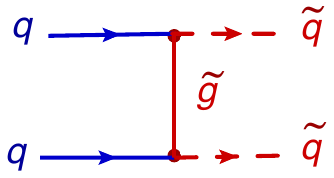}\\[4mm]
\includegraphics[width=0.24\textwidth,height=0.1\textwidth]{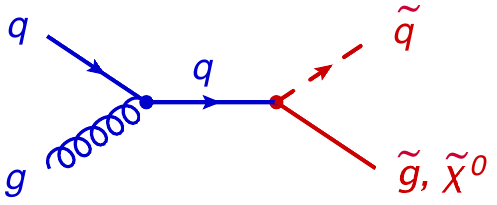}
\includegraphics[width=0.24\textwidth,height=0.1\textwidth]{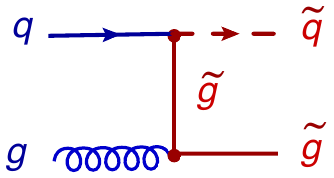}
\includegraphics[width=0.24\textwidth,height=0.1\textwidth]{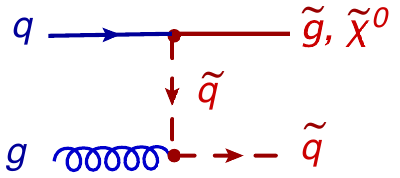}\\[4mm]
\includegraphics[width=0.24\textwidth,height=0.1\textwidth]{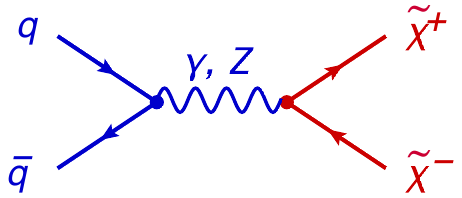}
\includegraphics[width=0.24\textwidth,height=0.1\textwidth]{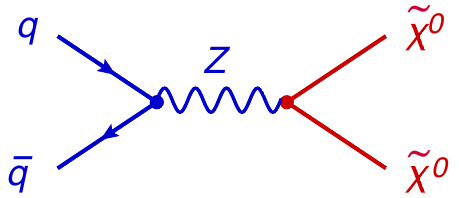}
\includegraphics[width=0.24\textwidth,height=0.1\textwidth]{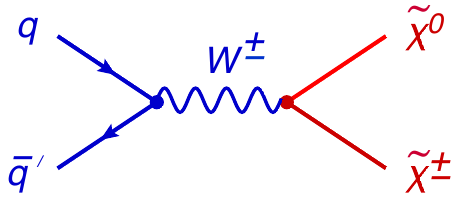}
\includegraphics[width=0.24\textwidth,height=0.1\textwidth]{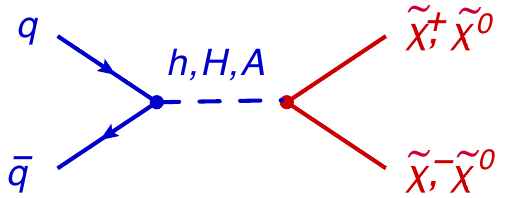} 
 \end{center}
 \caption{Examples of diagrams for SUSY particle production via strong interactions (top rows for  
 $\gl\gl$, $\sq\overline{\sq}$ and $\gl\sq$, respectively) and electroweak interactions (lowest row). }\label{f1}
 \end{figure}
\begin{figure}
 \begin{center}
 \includegraphics[width=0.45\textwidth]{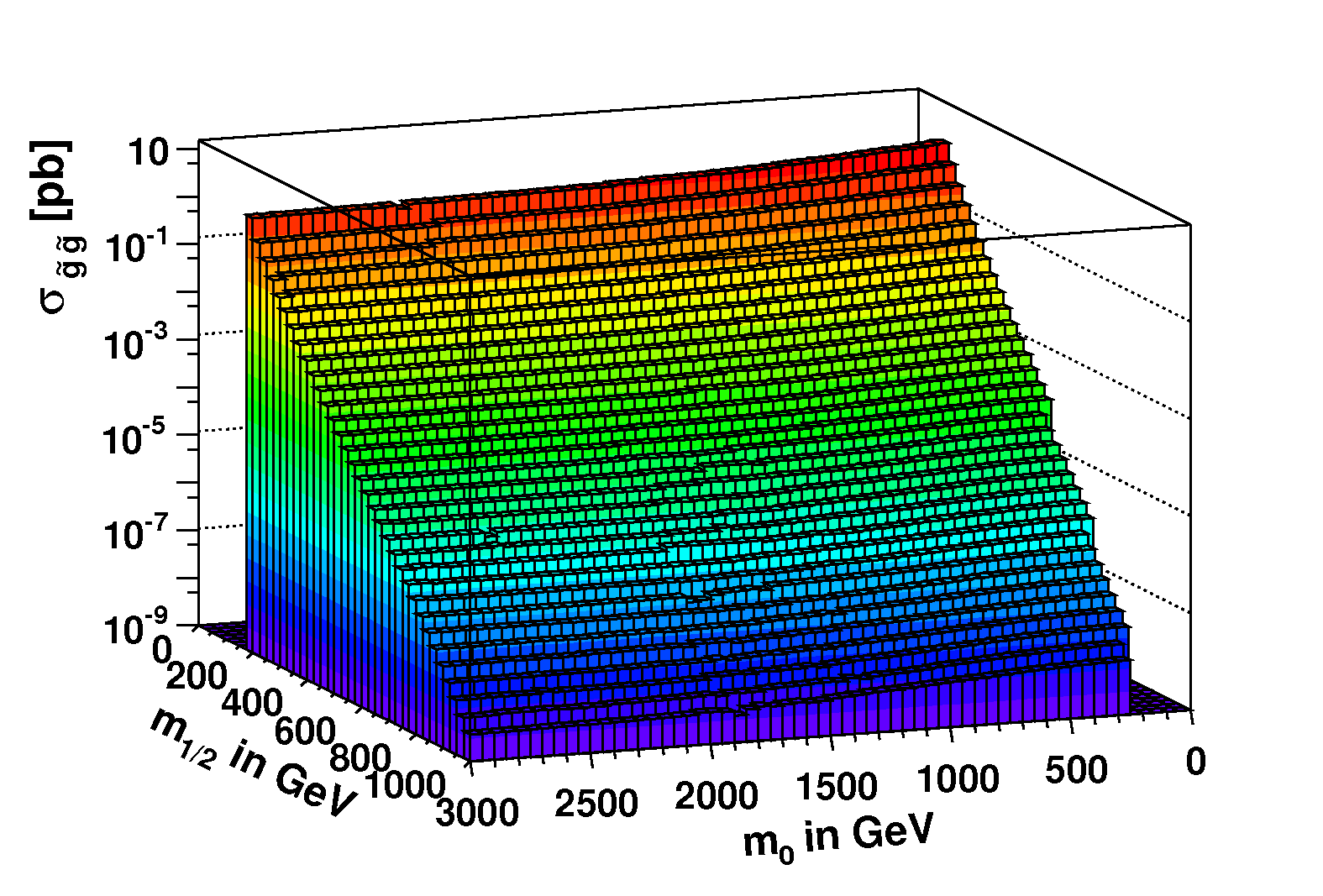}
\includegraphics[width=0.45\textwidth]{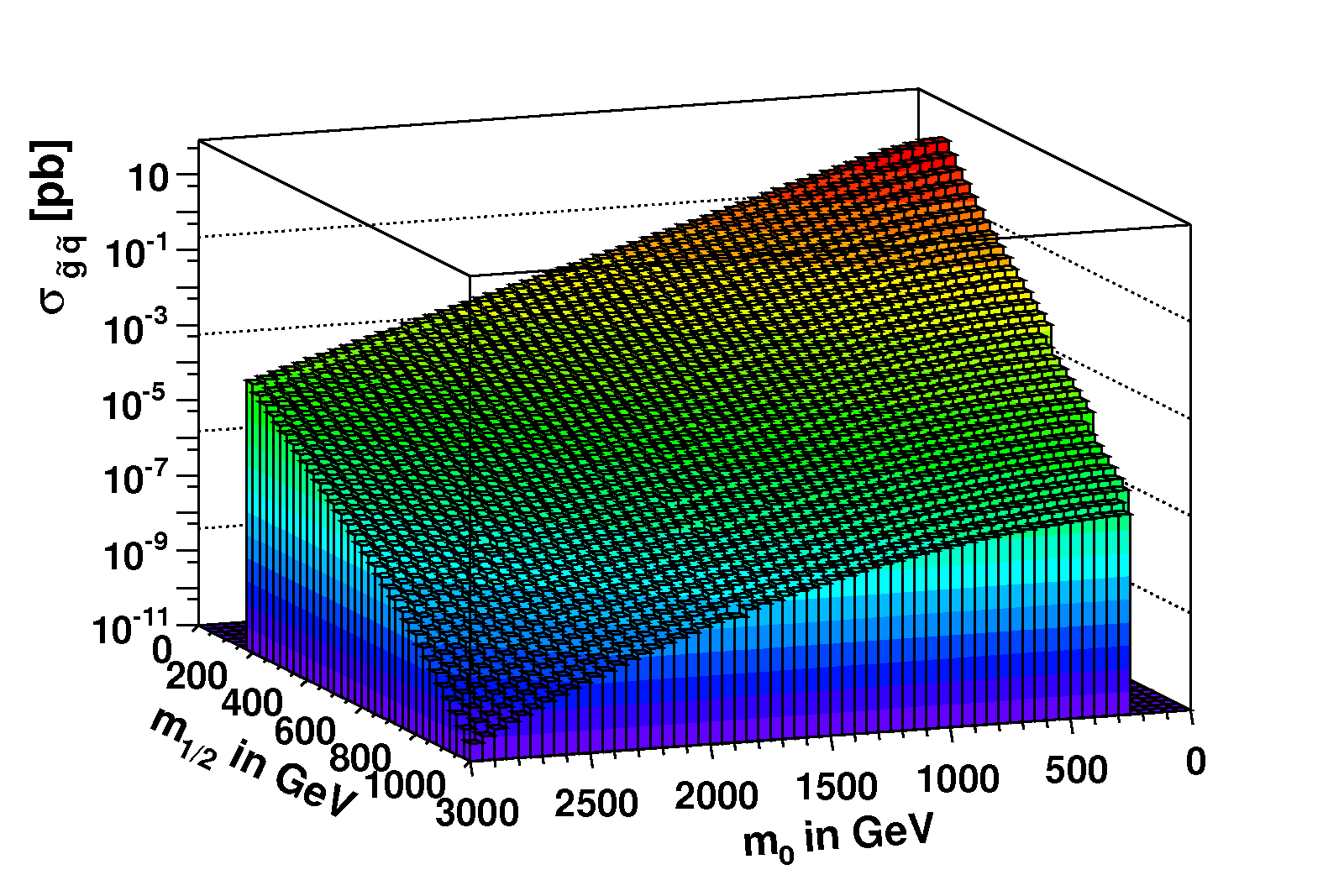}\\
\includegraphics[width=0.45\textwidth]{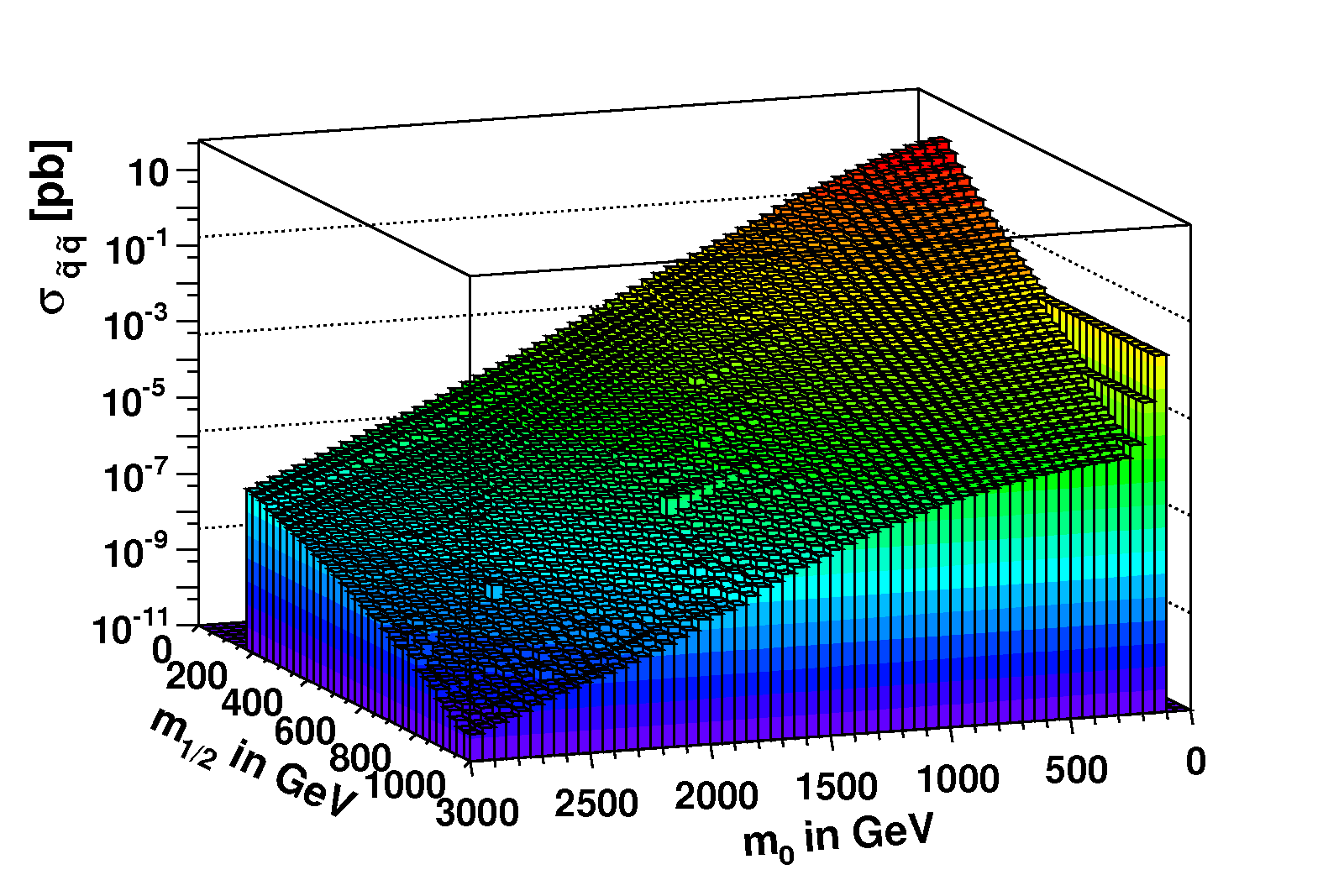}
\includegraphics[width=0.45\textwidth]{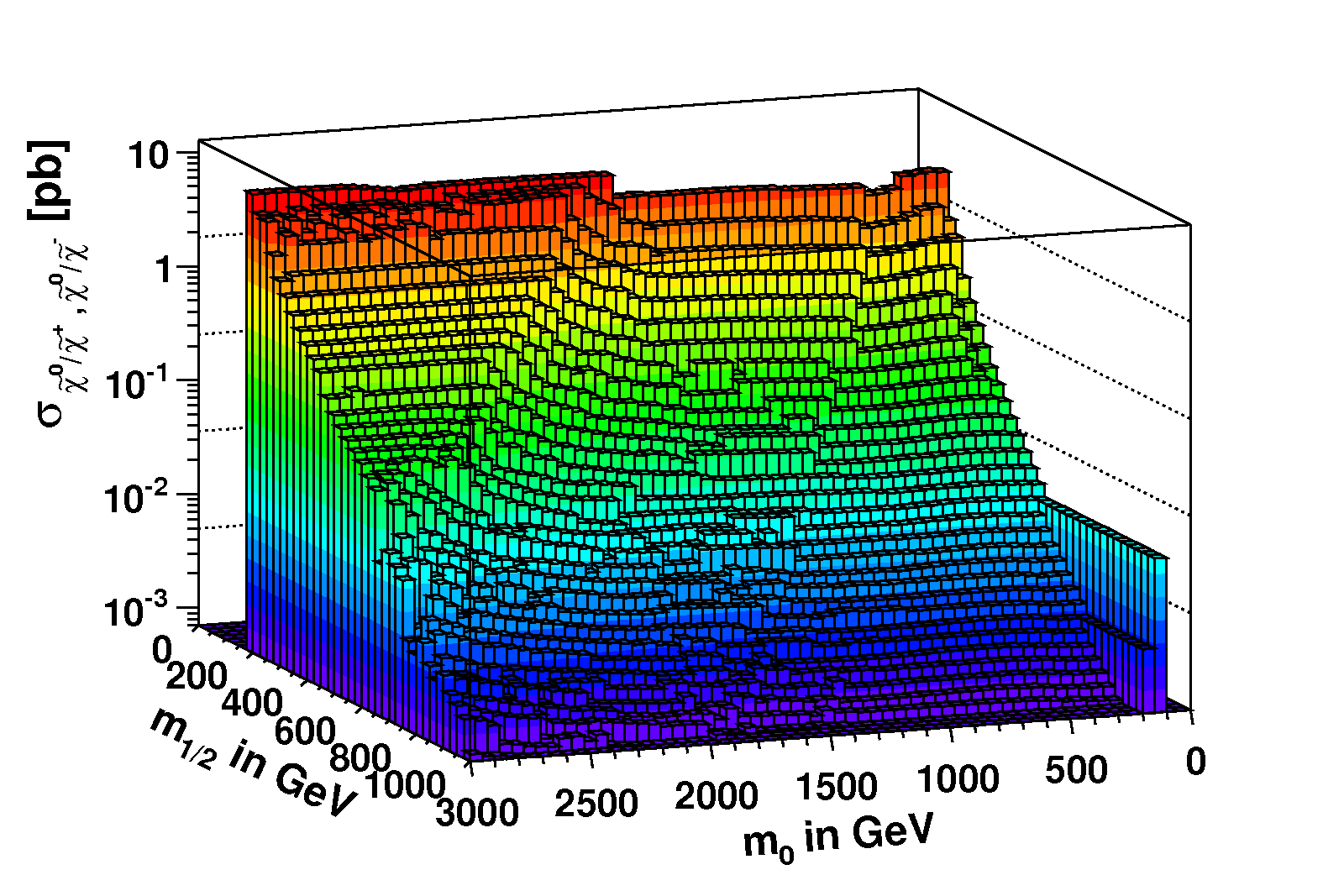}
 \end{center}%\vspace{-0.2cm}
 \caption{ Cross sections   for SUSY particle production for the diagrams shown in Fig. \ref{f1}: clockwise via strong interactions ($\gl\gl$,   $\gl\sq$ and  $\sq\overline{\sq}$, respectively) and electroweak interactions. }\label{f2}
 \end{figure} 
\begin{figure}
 \begin{center}
 \includegraphics[width=0.45\textwidth]{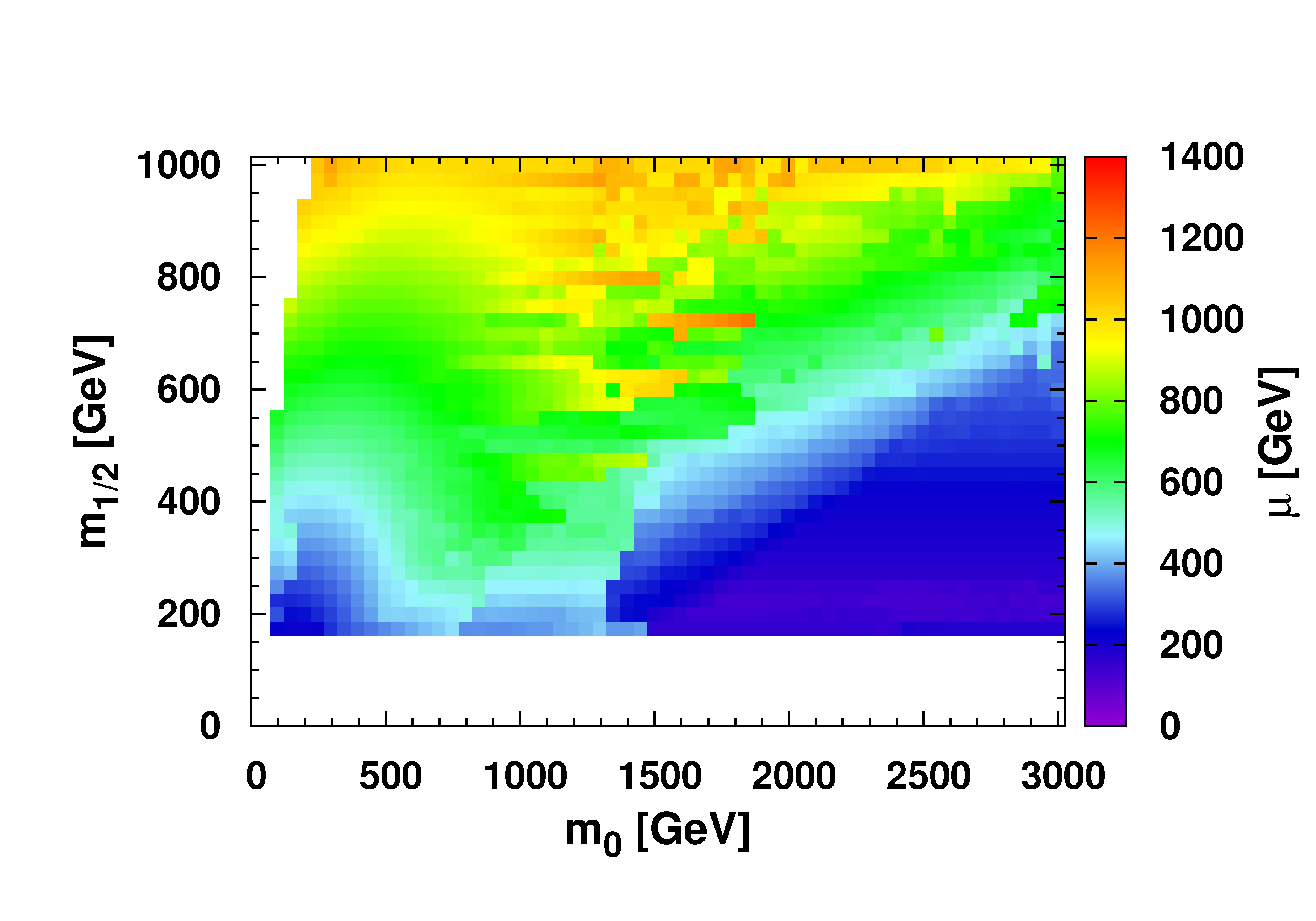}
\includegraphics[width=0.45\textwidth]{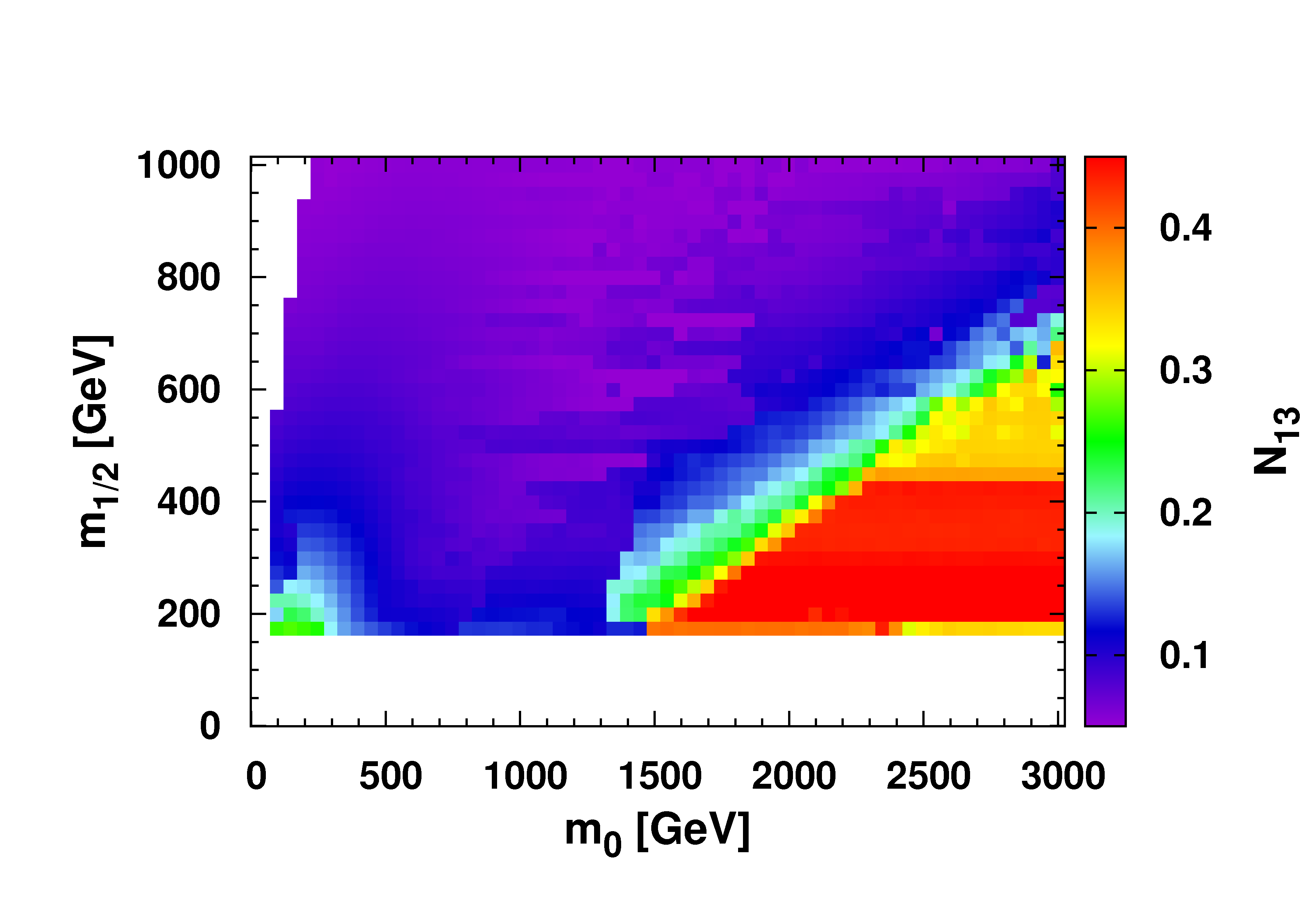}
 \end{center}%\vspace{-0.2cm}
 \caption{Values of $\mu$ and $N_{13}$  in the $\mzero,\mhalf$ plane. The strong decrease of $\mu$ and corresponding increase of $N_{13}$ at large values of $\mzero$ is largely determined by the EWSB constraint. Note that the right bottom corner is typically excluded by EWSB, {\it if} 
 the trilinear coupling  $A_0$ and $\tb$ are  fixed. However, if all parameters are left free 
 in the fit, EWSB can be fulfilled in the whole region shown. }\label{f3}
 \end{figure} 
\section{Excluded region by direct searches for SUSY at the LHC}\label{lhc}
In proton-proton collisions strongly interacting supersymmetric particles can be produced by the main diagrams shown in the first three rows of Fig. \ref{f1}, while the main  diagrams for  the electroweak production  are shown in the last row.
The corresponding cross sections are shown in Fig. \ref{f2} for a centre-of-mass energy of 7 TeV. One observes that the cross section for the
 ''strong'' production of  $\sq\overline{\sq}$ and  $\gl\sq$ )  are large for low values of $\mzero$ and $\mhalf$, the gluino production  $\gl\gl$ is strongest at small values of  $\mhalf$   and the electroweak production of gauginos starts to increase at large values of $\mzero$.
The reason for the increase of the electroweak production at large $\mzero$ is the decrease of the Higgs mixing parameter $\mu$, as determined from EWSB, which  leads to a stronger mixing of the Higgsino component in the gauginos and so the coupling to the weak gauge bosons and Higgs bosons increases, thus increasing the amplitudes for the diagrams in the last row of  Fig. \ref{f1}.

The physics is simple: the Higgs masses at the GUT scale start with a value of $\sqrt{\mzero^2 +\mu^2}$ and EWSB requires at least one of them to become negative at the electroweak scale by radiative corrections, mainly from the  Yukawa couplings of the third generation. This is only possible if the starting value is not too high, so a large value of $\mzero$ has to be compensated by a low value of $\mu$. This is demonstrated in Fig. \ref{f3}, which shows that $\mu$ becomes small and the Higgsino component in the lightest neutralino becomes large for large values of $\mzero$.

The strong production cross sections are characterized by a large number of jets from  long decay chains and missing energy from the escaping neutralino. These characteristics can be used to efficiently suppress the background. For the electroweak production, both the number of jets and the missing transverse energy is low, since the LSP is not boosted so strongly as in the decay of the heavier strongly interacting particles. Hence, the electroweak gaugino production needs leptonic decays to reduce the background, so these signatures need more luminosity and cannot compete at present with the sensitivity of the strong production of squarks and gluinos.

\begin{figure} 
 \begin{center}
 \includegraphics[width=0.45\textwidth]{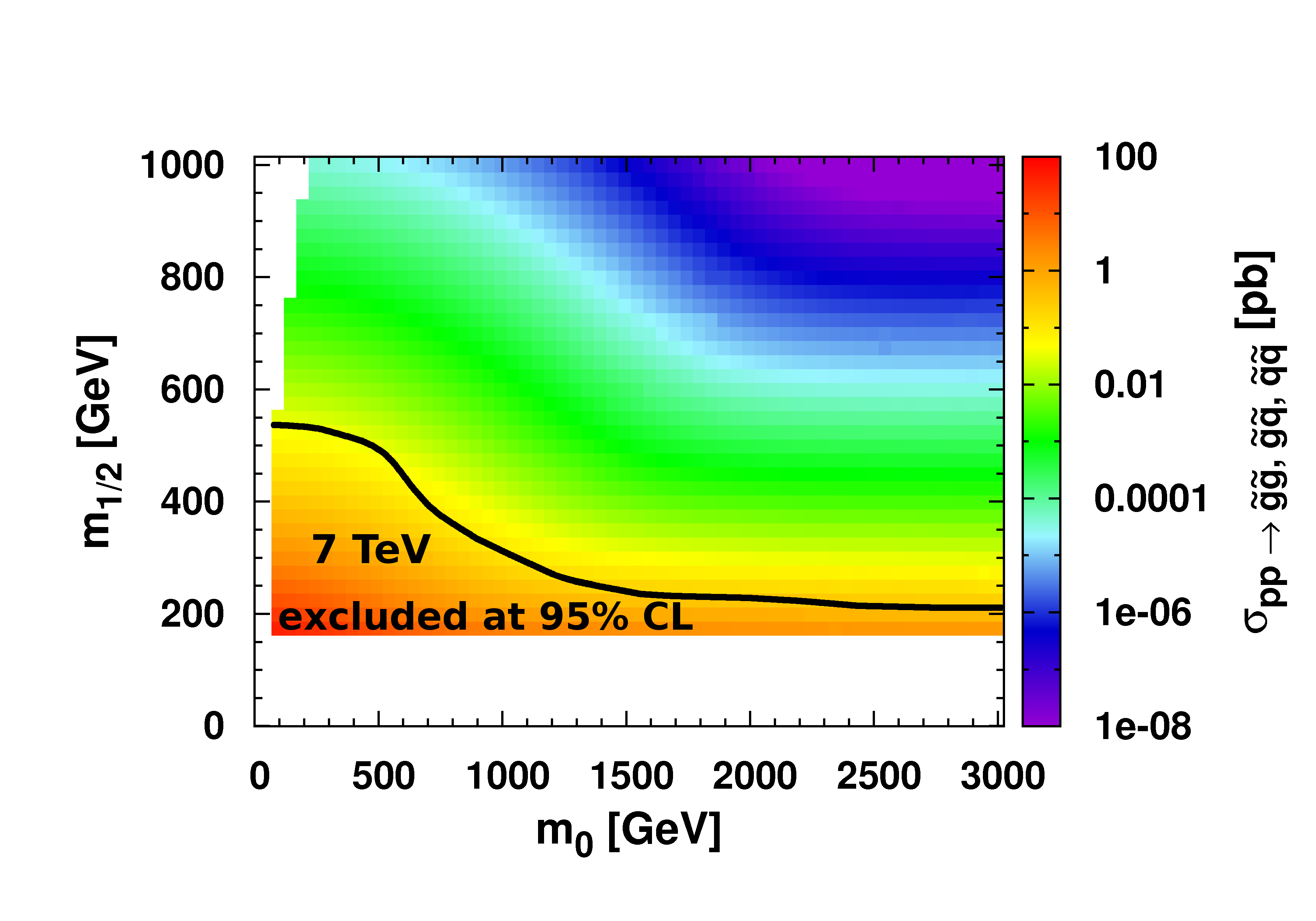}
\includegraphics[width=0.45\textwidth]{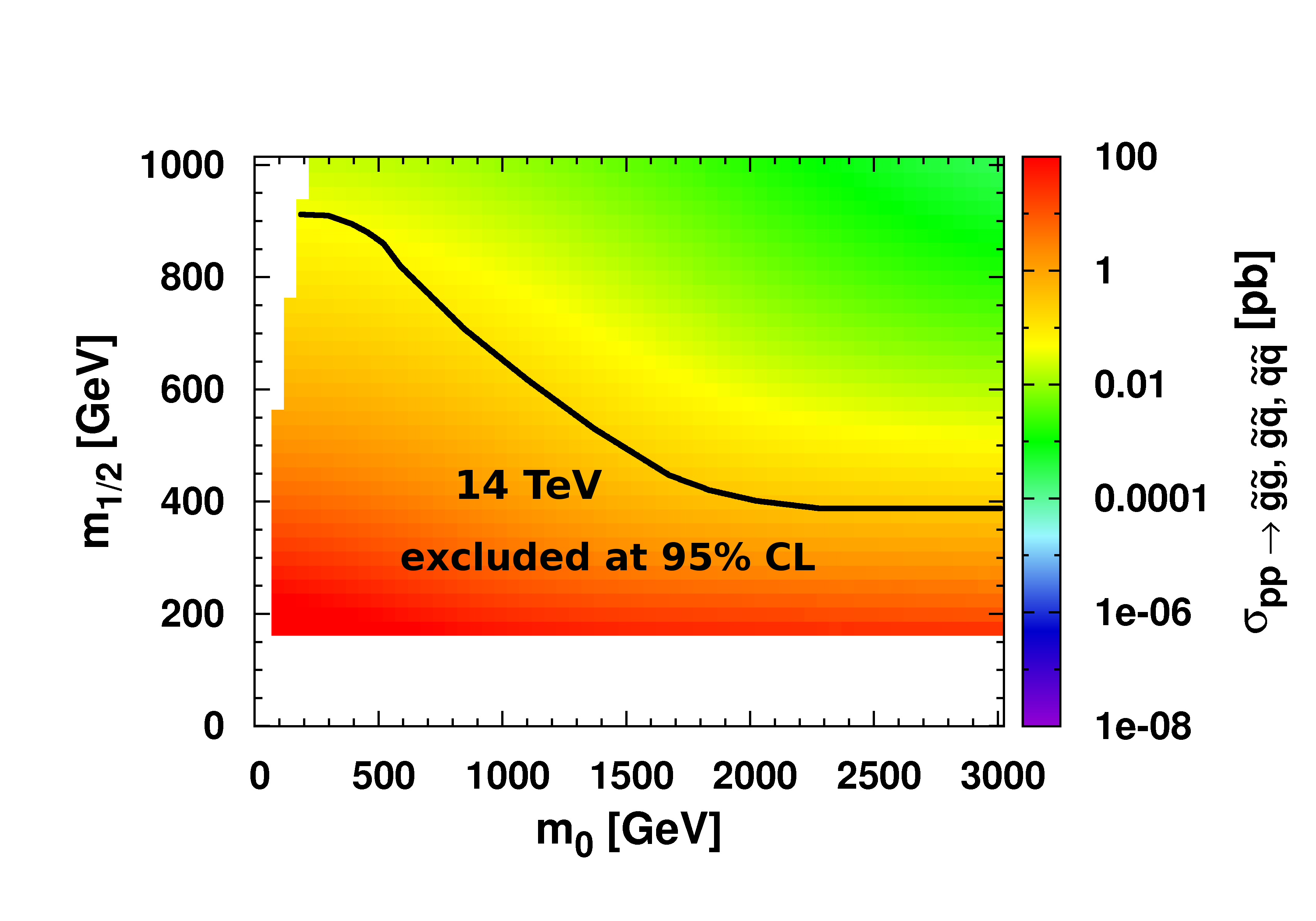}
 \end{center}%\vspace{-0.2cm}
 \caption{Left: Total production cross section of strongly interacting particles in comparison with the LHC excluded limits for 7 TeV. Here the data from ATLAS and CMS were combined and correspond to an integrated luminosity of 1.3 and 1.1 fb$^{-1}$, respectively. One observes that a cross section of 0.1 to 0.2 pb is excluded at 95\% confidence level. Right: the cross sections at 14 TeV and expected exclusion for the same limit on the cross section as at 7 TeV (left panel).
  }\label{f4}
 \end{figure}

The total cross section for the strongly interacting particles are shown in Fig. \ref{f4} together with the excluded region  from direct searches at the LHC for SUSY particles. One observes that the  excluded region (below the solid line) follows rather closely the total cross section, indicated by the colour shading. From the colour coding one observes that the excluded region corresponds to a cross section limit of about 0.1 -0.2 pb. The excluded region was obtained by combining the ATLAS \cite{Aad:2011qa} and CMS \cite{Chatrchyan:2011zy} limits in the following way. 
 Since the excluded region follows rather closely the total cross section the $\chi^2$ contribution has   been parametrized as $\sigma_{tot}^2/\sigma_{eff}^2$, where $\sigma_{eff}^2$ can be determined by requiring that for each point in the $\mzero,~\mhalf$ plane the $\chi^2$ value corresponds to an UP-parameter of $\chi^2-\chi_{min}^2$ = 5.99, which corresponds to a 95\% confidence level for a two-dimensional parameter space\footnote{This UP-parameter is for a two-sided confidence interval. However, several observables, like the LHC limits, refer to a single-sided lower limit.
  For a single-sided  limit the UP-parameter would be 4.61. 
 We keep conservatively an UP-parameter of 5.99, which for a single-sided limit would correspond to a 97.5\% confidence level Because of the steeply falling cross section the difference 
between 95 and 97.5\% is small for the exclusion line in the  $\mzero,~\mhalf$ plane.} \cite{James:1975dr}. Determining the  parameter $\sigma_{eff}^2$  for each point of the contour and each experiment independently implies that our contours are identical to the contours given by the experiments. Then the $\chi^2$ contributions are simply added, assuming no correlation between the experiments.   If the LHC data  is combined with cosmological and electroweak data, the fitted values of \tb~ and the trilinear coupling vary, while the LHC limits are given for fixed values $\tb=10$ and $A_0=0$. In our definition of $\chi^2$ the cross section variations as function of these parameters are correctly taken into account under the assumption that the efficiency does not vary with these parameters, which is a good approximation for the hadronic searches. Therefore we only consider limits from LHC data based on jets and missing energy and do not include the less sensitive limits from leptonic data. %Since the jet production cross section does not  depend strongly on $\tb$ and $A_0$ the corrections are anyway small and the procedure is accurate enough. 

The drop of the excluded region at large values of $\mzero$ is due to the fact that in this region the squarks become heavy, which means that the contributions from the diagrams in the second and third row of Fig. \ref{f1} start to diminish.  Here only higher energies will help and  
doubling  the LHC energy from 7 to 14 TeV, as planned in the coming years, quickly increases the cross section for gluino production by  orders of magnitude, as shown in the right panel of Fig. \ref{f4}. The expected sensitivity  at 14 TeV, plotted as an exclusion contour in case nothing will be found, assumes the same efficiency and luminosity (slightly above one fb$^{-1}$ per experiment) as at 7 TeV.
 
\begin{figure} 
 \begin{center}
 \includegraphics[width=0.45\textwidth]{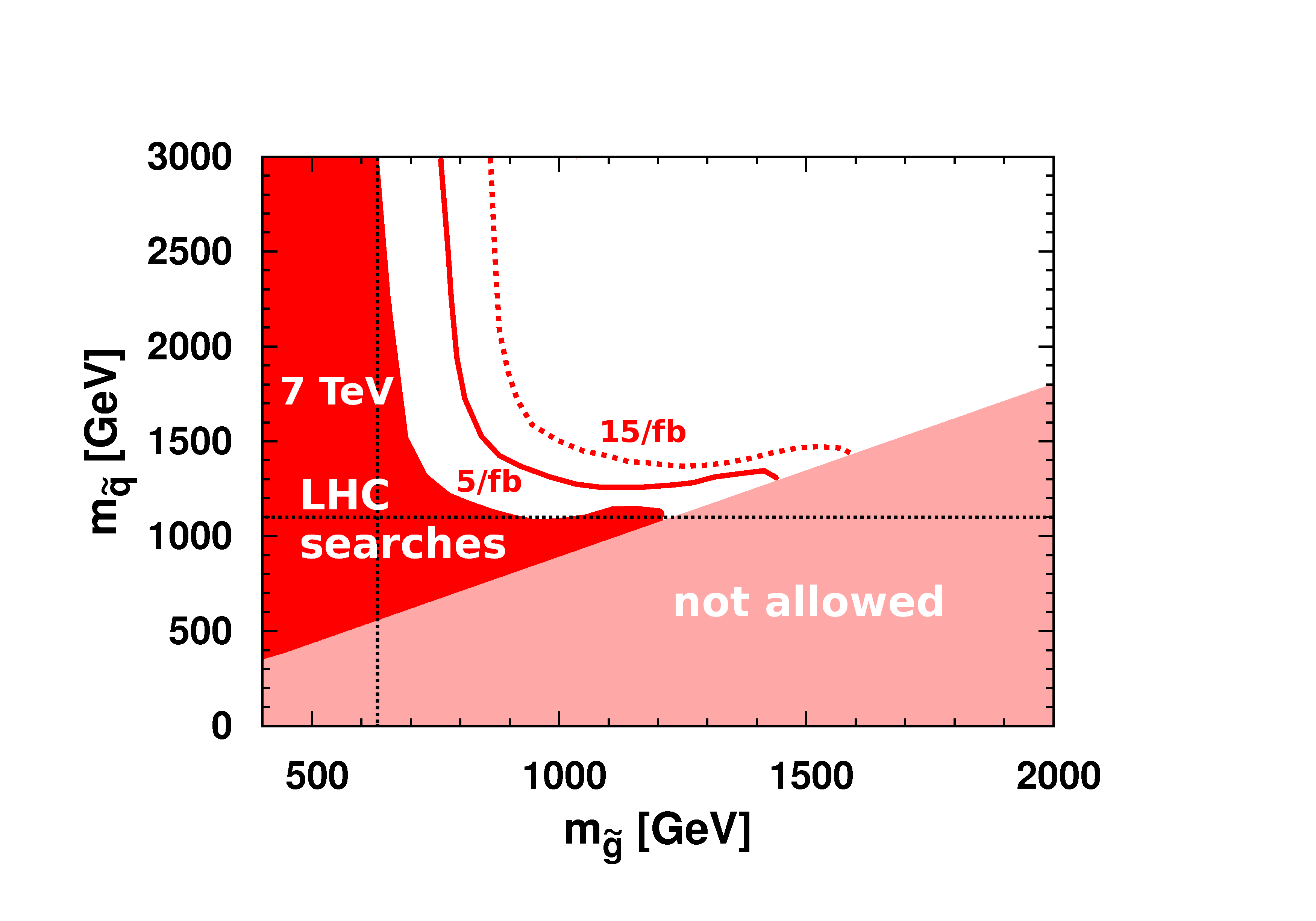}
\includegraphics[width=0.45\textwidth]{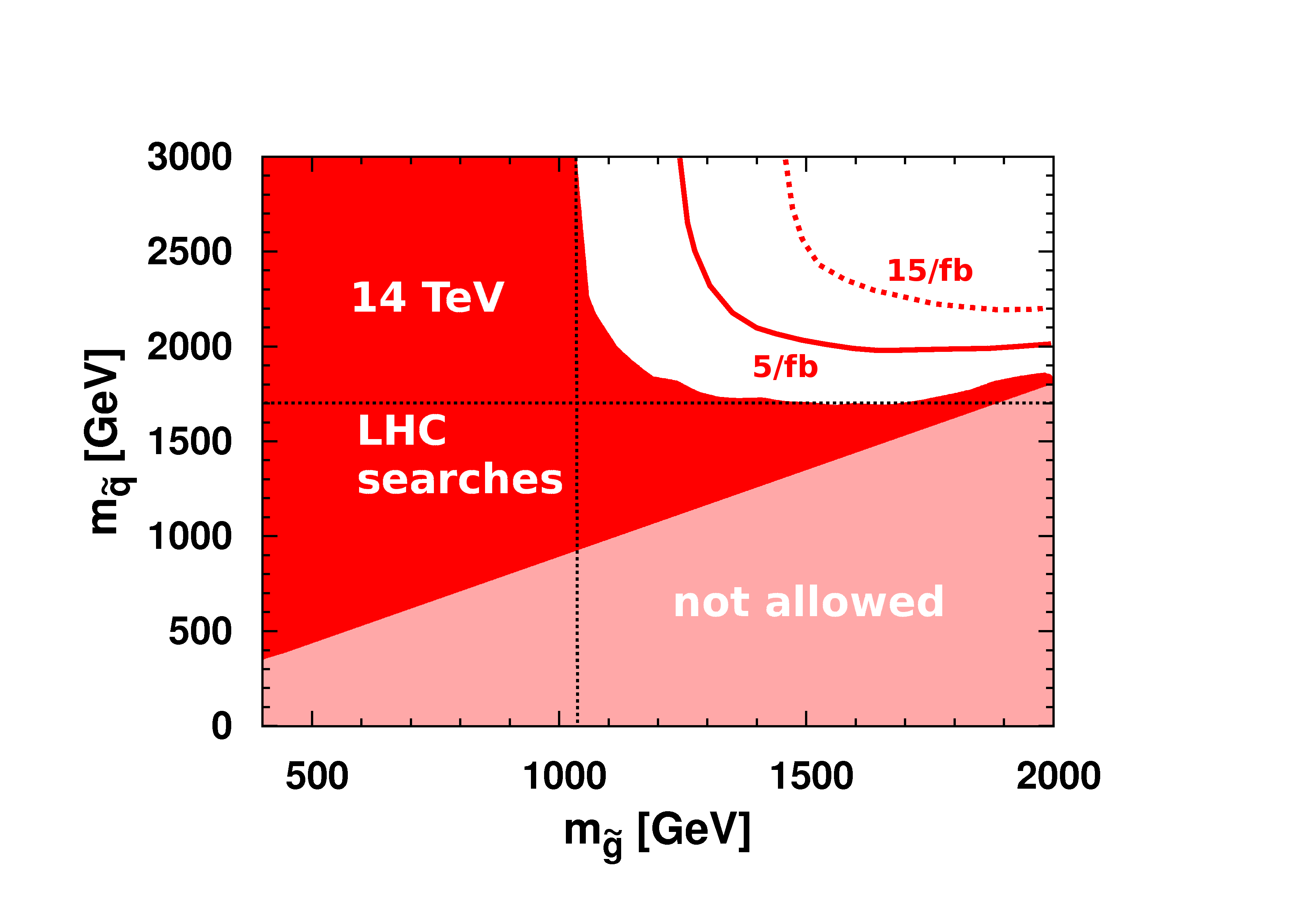}
 \end{center}%\vspace{-0.2cm}
 \caption{ As in Fig. \ref{f4}, but the excluded region is translated into the 
 $m_\sq ,m_\gl$ plane. The red area corresponds to excluded regions for an integrated luminosity slightly above 1/fb; the expectations for higher luminosities have been indicated as well. Note that the regions indicated as not allowed are not a feature of the CMSSM, but are valid in any model, which has a coupling between squarks and gluinos, which lead to self-energy diagrams of the squarks. These imply that if  the "gluino-cloud" is heavy, the squarks cannot be light.
  }\label{f5}
 \end{figure}

These limits can be translated to squark and gluino masses as follows. The squarks have a starting value at the GUT scale equal to $\mzero$, but have important contributions from gluinos in the colour field, so the squark masses are given by $m_{\sq}^2 \approx m_0^2+6.6\mhalf^2$, as was determined from the renormalization group equations \cite{deBoer:1994he}. Similarly the gluino mass is given by 2.7$\mhalf$. The  term proportional to $\mhalf$ in the squark mass corresponds to the self-energy diagrams, which imply that if the "gluino-cloud" is heavy, the squarks cannot be light. This leads to the regions indicated as not allowed in Fig. \ref{f5}. Note that these regions are forbidden in any model with a coupling between squarks and gluinos, so they are not specific to the CMSSM.
Squark masses below 1.1  TeV and gluino masses below 0.62  TeV  are excluded for the LHC data at 7  TeV, as shown in the left  panel of Fig. \ref{f5}.  Expected sensitivities for higher integrated luminosities at 7 and 14 TeV have been indicated as well.
One observes that increasing the energy is much more effective   than increasing the luminosity. At 14 TeV  squark masses of 1.7 TeV and gluino masses of 1.02  TeV are within reach with one fb$^{-1}$ per experiment, as shown in the right panel of Fig. \ref{f5}. 


\section{Excluded region by the relic density}\label{relic}
\begin{figure}
 \begin{center}
 \includegraphics[width=0.45\textwidth]{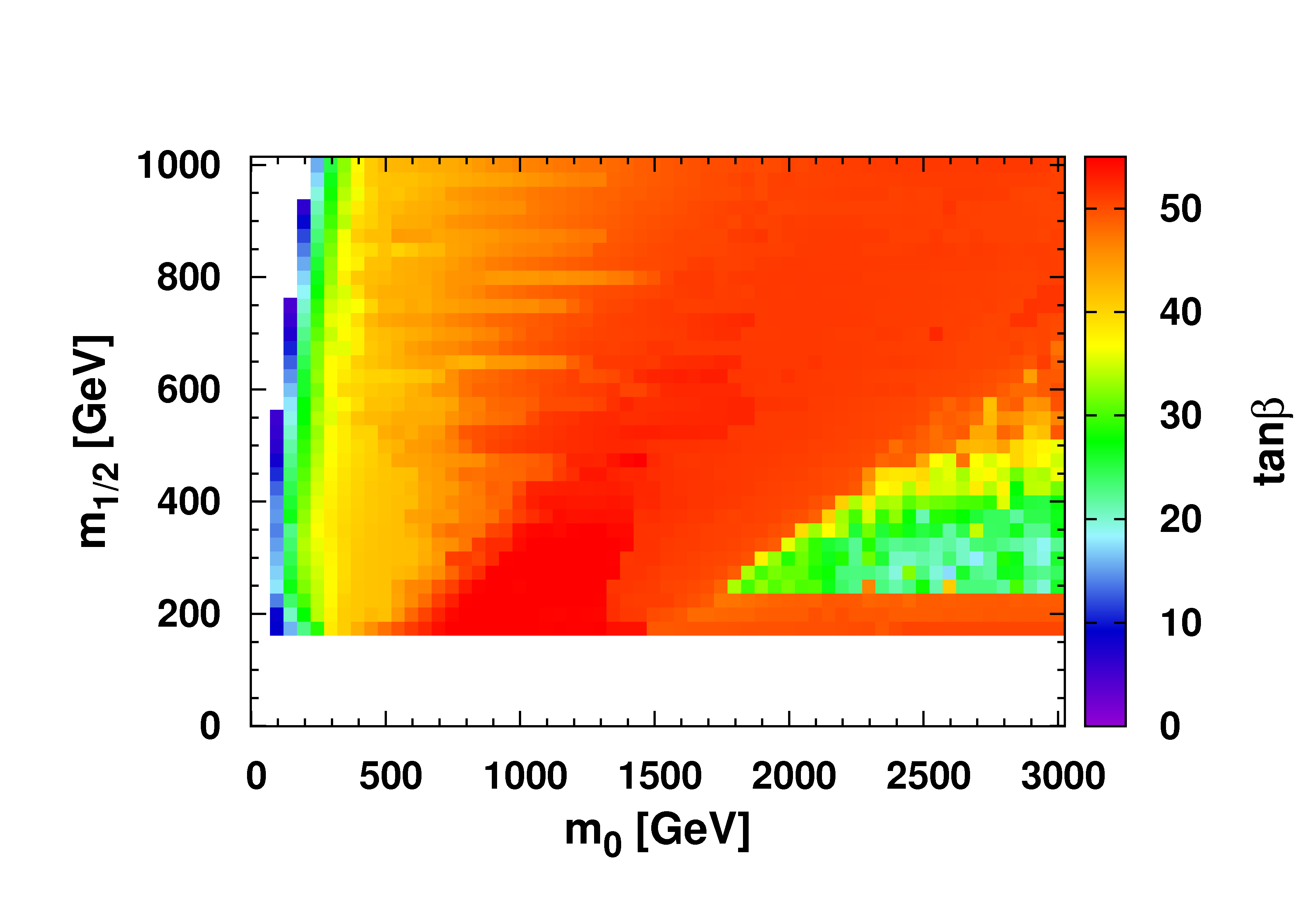}
\includegraphics[width=0.45\textwidth]{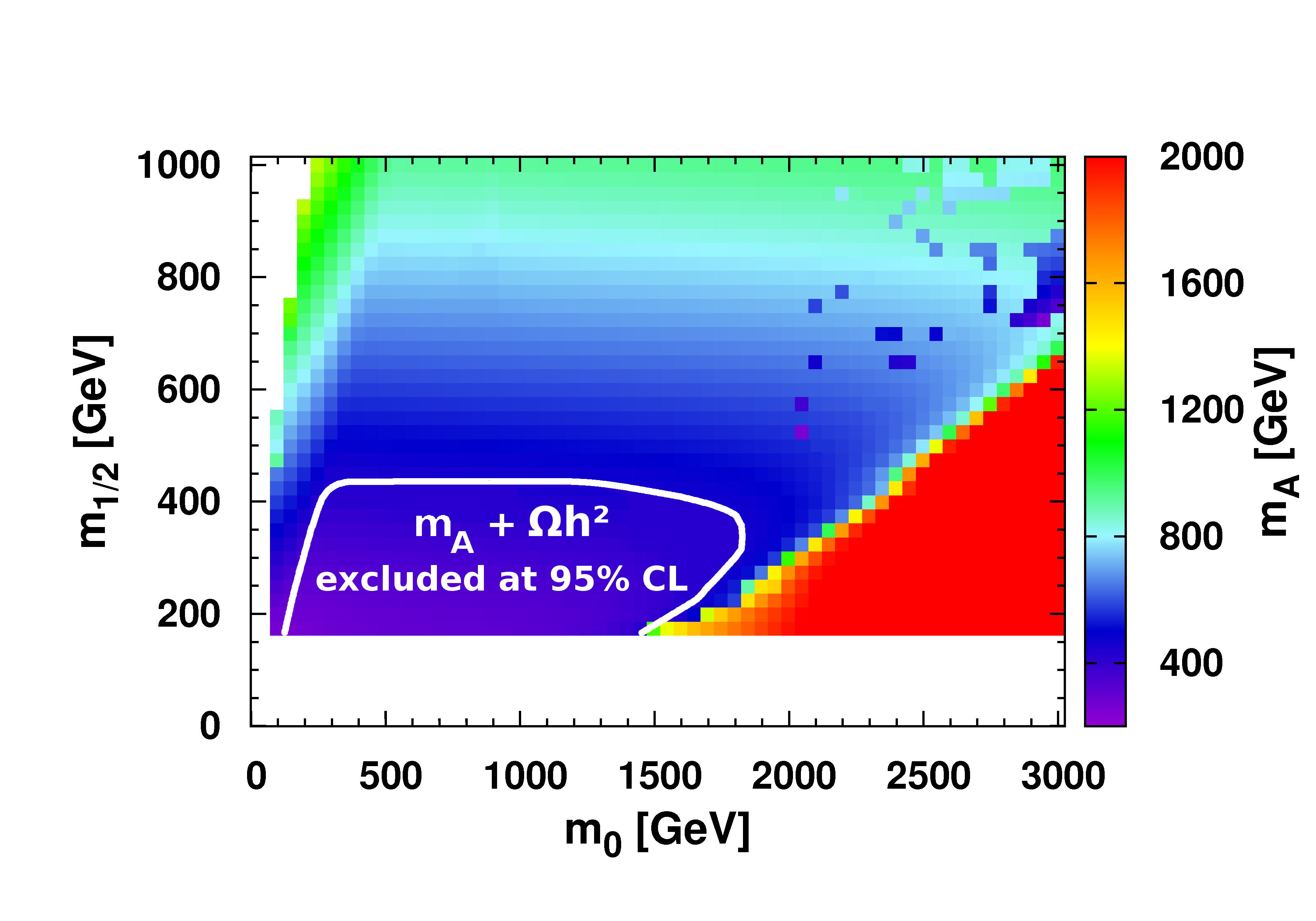}
 \end{center}%\vspace{-0.2cm}
 \caption{ Fitted values of $\tb$ (left) and constraint from relic density combined with the pseudo-scalar Higgs limit (right)  in the $\mzero,\mhalf$ plane after optimizing
 \tb~ and $A_0$ to fulfill the relic density and EWSB constraints at every point. The relic density requires $\tb\approx 50$ in most of the parameter space, where pseudo-scalar Higgs exchange dominates. In the (non-red) edges where $\tb$ is lower, co-annihilation  contributes.
 The data below the solid line in the right panel is excluded at 95 \% confidence level.
}\label{f6}
 \end{figure}

The observed relic density of DM corresponds to $\Omega h^2=0.113\pm 0.004$ \cite{Komatsu:2010fb},  which is about a factor six higher than the baryonic density. This number is inversely proportional to the annihilation cross section.
The dominant annihilation contribution comes from A-boson exchange in most of the parameter space, if one excludes the narrow co-annihilation regions \cite{Beskidt:2010va}. The cross section for $\chi+\chi \to A \to b\bar b$
 can be written as:
\beq <\sigma v>\sim  \frac{M_\chi^4
m_b^2 \tan^2\beta}{\sin^4 2\theta_W
\,M_Z^2}\frac{ \left( N_{31}\sin\beta -N_{41}\cos\beta
\right)^2\left( N_{21}\cos\theta_W - N_{11}\sin\theta_W
\right)^2}{\left( 4M_\chi^2 - M_A^2
\right)^2+M_A^2\Gamma_A^2},
\eeq
where the elements of the mixing matrix in the neutralino sector define the content of the lightest neutralino
\beq
|\tilde \chi^0_1\rangle =N_{11}|B_0\rangle
+N_{12}|W^3_0\rangle +N_{13}|H_1\rangle +N_{14}|H_2\rangle .
\eeq
 The correct relic density requires $<\sigma v>=2\cdot  10^{-26}\ cm^3/s$, which implies that the annihilation cross section $\sigma$ is of the order of a few pb. Such a high cross section can be obtained only close to the resonance. Actually on the resonance the cross section is too high, so one needs to be in the tail of the resonance, i.e. $m_A\approx 2.2m_\chi$ or $m_A\approx 1.8 m_\chi$. So one expects $m_A \propto m_{1/2}$ from the relic density constraint. This constraint can be fulfilled with \tb~ values around 50 in the whole $\mzero,\mhalf$ plane, except for the narrow co-annihilation regions, as we showed previously \cite{Beskidt:2010va}. The results were extended to larger values of $\mzero$, as shown in the left panel of Fig. \ref{f6}.  The rather strong limits on the pseudo-scalar Higgs boson from LHC \cite{Chatrchyan:2011nx,Aad:2011rv}, especially at large values of \tb, lead then to constraints on $\mhalf$ of about 400 GeV for intermediate values of $\mzero$, as shown in the right panel of Fig. \ref{f6}. Note that the top left corner (white) is not allowed, because in this region of small
 $\mzero$ and large $\mhalf$ the stau becomes the lightest supersymmetric particle (LSP).
 However, a charged LSP cannot be the DM. As will be shown later in Fig. \ref{f10}, for even larger values 
 of $\mhalf$ this non-allowed region becomes broader, while the relic density requires large $\tb$
 away from the co-annihilation region, in which case the mixing in the stau sector increases and the
 stau becomes again the LSP. 
%The corresponding values of \tb\ needed are shown in the right panel of Fig. \ref{f2}. The production cross section of the pseudo-scalar Higgs at the LHC is proportional to $\tan^2\beta$, so the present limits from LHC are proportional to  $\tb^2$ as well. At \tb =50 the present limit on $m_A$ is about 450 GeV, so the limit on $m_{1/2}$ is close to it. The exclusion on $m_A$  as function of \tb\ from the CMS collaboration \cite{cmsma} is indicated in Fig. \ref{f4} as well. Similar limits have been obtained by the ATLAS collaboration \cite{Aad:2011rv}.  
\begin{figure}
 \begin{center}
 \includegraphics[width=0.5\textwidth,height=0.15\textwidth]{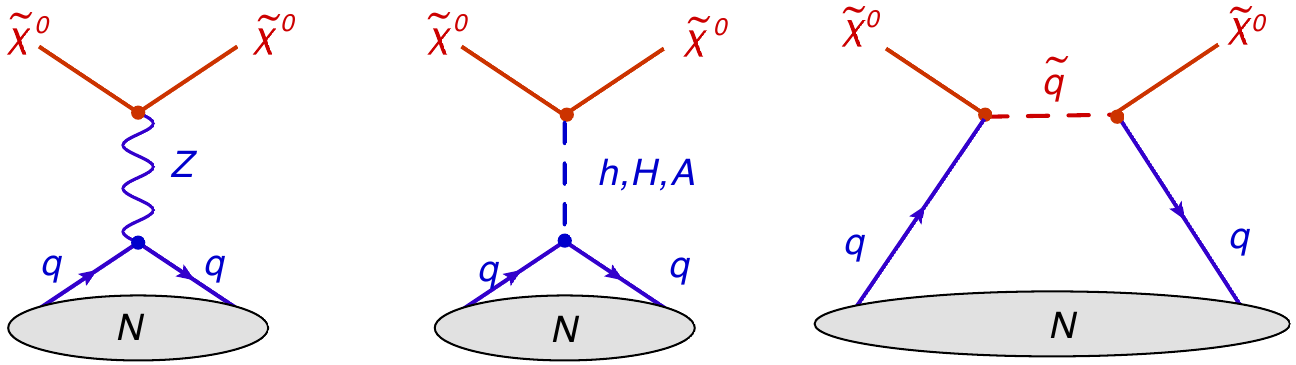}
 \end{center}
 \caption{Examples of diagrams for elastic neutralino-nucleon scattering. }\label{f7}
 \end{figure}

 \section{Excluded region by direct DM searches}\label{direct}
Scattering of LSPs on nuclei can only happen via elastic scattering, if
R-parity is conserved \cite{Jungman:1995df,Kolb:1990vq}. The corresponding diagrams are shown in
Fig. \ref{f7}.    The big blob
indicates that one enters a low energy regime, in which case the protons and neutrons inside the nucleus cannot be resolved.
In this case the spin independent scattering becomes coherent on all nuclei and the cross section becomes proportional to the number of nuclei:
\begin{equation}
\sigma=\frac{4}{\pi}\frac{m_{\rm DM}^2m_N^2}{(m_{\rm DM}+m_N)^2}\left(Z
f_p+(A-Z)f_n\right)^2
\label{sc}
\end{equation}
where $A$ and $Z$ are the atomic mass and atomic number of the target
nuclei. 
  Since the particle
which mediates the scattering is typically much heavier than the momentum
transfer, the scattering can be written in terms of an
effective coupling. Using the notation of Ref. \cite{Ellis:2008hf} one can write:
\begin{equation} \label{fpn}
f_{p,n}=\sum_{q=u,d,s} G_{q}
f^{(p,n)}_{Tq}\frac{m_{p,n}}{m_q}+\frac{2}{27}f^{(p,n)}_{TG}\sum_{q=c,b,t}
G_{q}\frac{m_{p,n}}{m_q},
\end{equation}
where $G_q=\lambda_{\rm DM}\lambda_q / M^2_M$. Here $M$ denotes the
mediator, and $\lambda_{\rm DM}$, $\lambda_{f}$ denote the mediator's
couplings to DM and quark. The parameters $ f^{(p)}_{Tq}$
are defined by
\begin{equation} m_p  f^{(p)}_{Tq} \equiv \langle p|m_q\bar q q|p \rangle \label{condensate}
\end{equation}
and similar for $ f^{(n)}_{Tq}$,
whilst $ f^{(p,n)}_{TG}=1-\sum_{q=u,d,s} f^{(p,n)}_{Tq}$.

When squarks become heavy, the squark exchange diagram in Fig. \ref{f7} becomes   suppressed and the only relevant mediator is the Higgs boson, in which case the  heavier sea quarks
in the sum in Eq. \ref{fpn} become the main contributor to the elastic cross section.
Remember that the coupling between the Higgs and down-type quarks is additionally enhanced by $\tb$ \cite{Haber:1984rc,Gunion:1984yn}, so the charm and top contributions are suppressed for the large values of $\tb$ in Fig. \ref{f6}.
% since the couplings of the light quarks inside the protons are suppressed by the small mass, while the  heavier quarks are suppressed by the scalar density of the heavy quarks inside a proton or neutron.

The scalar density  of the heavy quarks can be inferred from the nucleon masses, since a large 
density of heavy quarks would increase the nucleon mass.
The nucleon mass can be written as $M_N=M_0+\sigma_{\pi N}$, where $M_0$ is the nucleon mass  in the chiral limit and  $\sigma_{\pi N}$ characterizes the effect of the finite quark masses. The $u,d$ quarks contribute both as valence and sea quarks, while the heavier quarks contribute only as sea quarks.
%The quark density is given by the expectation of the number of $q\overline{}$ pairs in the nucleon, which can symbolically be written as the vacuum subtracted matrix element $\angle N  | q\overline{q} | N\rangle$ and $\sigma_{\pi N}= \sum_q m_q \langle N| q\overline{q}| N\rangle$. 
The strange quark content is usually parametrized by the $y$-parameter: $$y=\frac{2\langle N| s\overline{s}| N\rangle}{\langle N| u\overline{u} + d\overline{d}| N\rangle}.$$
Phenomenologically, the $\pi N$ scattering is related to the $\sigma_{\pi N}$ term, while the contribution of heavier quarks is related to the octet breaking of the baryon masses. From these data one obtains $y$=0.3-0.6 \cite{Ohki:2008ff}. Such large values are surprising, but one should keep in mind that they originate from the gluon splitting inside a proton, which is the non-perturbative regime of QCD and the interpretations might be prone to higher order corrections.

An alternative solution for calculating the scalar quark densities in the non-perturbative regime is provided by lattice QCD. Such calculations allow to determine the valence and sea quark distributions to the $\sigma_{\pi N}$ independently. After imposing the chiral symmetry in the lattice calculations the authors of Ref. \cite{Ohki:2008ff} find that the valence quarks dominate the $\sigma_{\pi N}$ term and the $y$ parameter is small $(< 0.05)$.  

%The cross sections can be parametrized by effective couplings $f^{(p)}_{Tq}$, which are related to the nuclear matrix elements, e.g. $f^{(p)}_{Ts}={m_s\langle N\mid s\overline{s}\mid N\rangle}/M_N$.
 The default values of the effective couplings in Eq. \ref{condensate} are  in micrOMEGAs \cite{Belanger:2008sj}: $f^{(p)}_{Tu}=0.033,~f^{(p)}_{Td} =
0.023,~f^{(p)}_{Ts}=0.26,~f^{(n)}_{Tu}=0.042,~f^{(n)}_{Td}=0.018,~f^{(n)}_{Ts}=0.26$.
%More recent values of the $\pi N$ sigma term yield 68 MeV instead of 45 MeV, which increases the scalar density of the  strange quarks from $y=0.2\pm0.2$ to $y=0.47\pm0.2$ and $f^{(p)}_{Ts}$ from 0.118 to 0.41  \cite{Ellis:2008hf}. 
%TODO EXPLAIN LATTICE BETTER from ohki or buchmueller
If one takes  the lower $y$ values from the  lattice calculations one finds \cite{Cao:2010ph}:
$f^{(p)}_{Tu}=0.020,~f^{(p)}_{Td} =
0.026,~f^{(p)}_{Ts}=0.02,~f^{(n)}_{Tu}=0.014,~f^{(n)}_{Td}=0.036,~f^{(n)}_{Ts}=0.02$.
So the most important coupling to the strange quarks is reduced from 0.26 to 0.02,  which implies an order of magnitude  uncertainty in the elastic neutralino-nucleon scattering cross section.

%Also the b-quark contribution becomes non-negligible \cite{Giedt:2009mr}.
%%%%%%%%%%%%%%%%%%
Another normalization uncertainty in direct dark matter experiments arises from
the uncertainty in the local DM density, which
can take values between 0.3 and 1.3 GeV/cm$^3$, as determined from the
rotation curve of the Milky Way, see Ref. \cite{Weber:2009pt,deBoer:2010eh} and references
therein.

These overall normalization uncertainties from the scalar quarks densities inside the nucleon and the DM densities in the Galaxy are independent of the SUSY parameter space. However, the effective couplings vary strongly inside the SUSY parameter space.
%In the limit of zero momentum transfer  the effective couplings take the
%form:
%\begin{eqnarray*}
% G_q(A)& =&0, \label{sc1} \\
%G_u(h)&=& \frac{-e^2 m_u}{2\sin^2 2\theta_W M_Z} \left(\!
%N_{21}\cos\theta_W\! - \!N_{11}\sin\theta_W\! \right)
%\frac{\cos\alpha}{\sin\beta}
%  \frac{ \left(\! N_{41}\cos\alpha\! + \!N_{31}\sin\alpha\!
%\right)}{M_h^2}, \\
%G_d(h)&=& \frac{e^2 m_d}{2\sin^2 2\theta_W M_Z} \left(\!
%N_{21}\cos\theta_W\! -\! N_{11}\sin\theta_W\! \right)
%\frac{\sin\alpha}{\cos\beta}
%  \frac{ \left(\! N_{41}\cos\alpha\! +\! N_{31}\sin\alpha\!
%\right)}{M_h^2}, \\
%G_u(H)&=& \frac{-e^2m_u}{2\sin^2 2\theta_W M_Z} \left(\!
%N_{21}\cos\theta_W\! -\! N_{11}\sin\theta_W \!\right)
%\frac{\sin\alpha}{\sin\beta}
%   \frac{\left(\! N_{41}\sin\alpha\! -\!N_{31}\cos\alpha\!
%\right)}{M_H^2}.\\
%G_d(H)&=& \frac{-e^2m_d}{2\sin^2 2\theta_W M_Z} \left(\!
%N_{21}\cos\theta_W\! -\! N_{11}\sin\theta_W \!\right)
%\frac{\cos\alpha}{\cos\beta}
%   \frac{\left(\! N_{41}\sin\alpha\! -\!N_{31}\cos\alpha\!
%\right)}{M_H^2}.
%\end{eqnarray*}

% The Higgs mixing term $\mu$, which is determined by EWSB, becomes small at
%large values of $m_0$ and the Higgsino components  $N_{31},N_{41}$ become
%correspondingly large, as shown in Fig. \ref{f3}.
 The scattering cross section is proportional to the
product of gaugino and Higgsino components squared \cite{Haber:1984rc,Gunion:1984yn}, so it rises rapidly, if the Higgsino component increases, which is the case for large values of $\mzero$,  as shown before in Fig. \ref{f3}.
  To get conservative estimates for the excluded regions, we take the
lowest possible values of the local DM density and the couplings.
 The difference in couplings is the largest contributor to the
uncertainty and the difference in excluded regions is demonstrated in Fig.
\ref{f8} (left panel).
The dotted line  corresponds to the default micrOMEGAs values of the couplings based on the $\sigma_{\pi N}$  term from data, while the dashed-dotted line indicates the 95\% excluded region using the smaller and hence more conservative  couplings from lattice gauge theory. The colours indicate the $\Delta \chi^2=\chi^2-\chi_{min}^2$ contribution from the electroweak constraints: $\bsg=(3.55 \pm 0.24)\cdot 10^{-4}$ \cite{hfag}, g-2 \cite{Bennett:2006fi} , $\bsmm < 1.1\cdot 10^{-8}$ \cite{cmslhcb}, $\btaunu =(1.68\pm 0.31)\cdot 10^{-4}$ \cite{hfag}, $m_h>114.4$ GeV \cite{Schael:2006cr}. $\chi_{min}=2.5$ in this plot for two degrees of freedom, if we consider only the
measured values (excluding limits) minus the two fitted SUSY parameters $A_0$ and \tb. 

If we combine all constraints one finds that $\mhalf$ below 400 GeV is excluded for all values of $\mzero$, as shown
in the right panel of Fig. \ref{f8}.
At low values of $\mzero$ the direct searches dominate the limit, as shown previously in Fig. \ref{f4}. 
At intermediate values of $\mzero$ the relic density constraint dominates the limit, as shown  in Fig. \ref{f6}, while 
at large values the limit is dominated by the XENON100 data (Fig. \ref{f8},left panel).

It should be noted that these combined limits from the LHC, WMAP and XENON100 are  stronger
than the limits from the LEP Higgs limits, electroweak and b-physics observables (red area in Fig. \ref{f8}, left), so these hardly play a role anymore. Only g-2 yields a mild, but insignificant preference for low values of $\mzero$, especially if one combines the non-Gaussian theoretical errors linearly with the experimental errors. This will be discussed in more detail below.

\subsection{Discussion on g-2}
 The theoretical value of g-2 has been reviewed in Ref. \cite{Jegerlehner:2009ry}, which is still in agreement  with the latest values from Ref. \cite{Davier:2010nc}.
 We use the difference $\Delta a_\mu$ in g-2 between the experimental and theoretical value as input to the $\chi^2$ value and use $\Delta a_\mu=302~\pm~63(exp)~\pm~61(theo.)\cdot 10^{-11}$, so the theoretical and experimental errors are similar. The theoretical error from the light-by-light scattering contribution alone is 26, so this error dominates, but it has certainly not a Gaussian distribution. For non-Gaussian errors, especially errors with an equal probability in a certain interval, it is more conservative to add the errors linearly, as can be tested by simply comparing the convolution of two Gaussians with a Gaussian and a "flat top" Gaussian, where the flat region represents the interval with a constant probability. In the first case one finds the usual result that the probability distribution from the combined uncertainties equals  a Gaussian with the errors added in quadrature, while in the second case the probability distribution equals a Gaussian with an error closer to the linear addition of the individual errors.  
To be conservative we have added the theoretical and experimental errors linearly, which increases the error for g-2 from 88 to 124 in units of $10^{-11}$, so the $\chi^2$ contribution is reduced by about a factor two, but there is still a preferred region with lower
$\chi^2$, as shown in Fig. \ref{f9}, left panel. To be consistent we have added  theoretical
and experimental errors linearly for all other observables as well. 

\begin{figure}
 \begin{center}
 \includegraphics[width=0.45\textwidth]{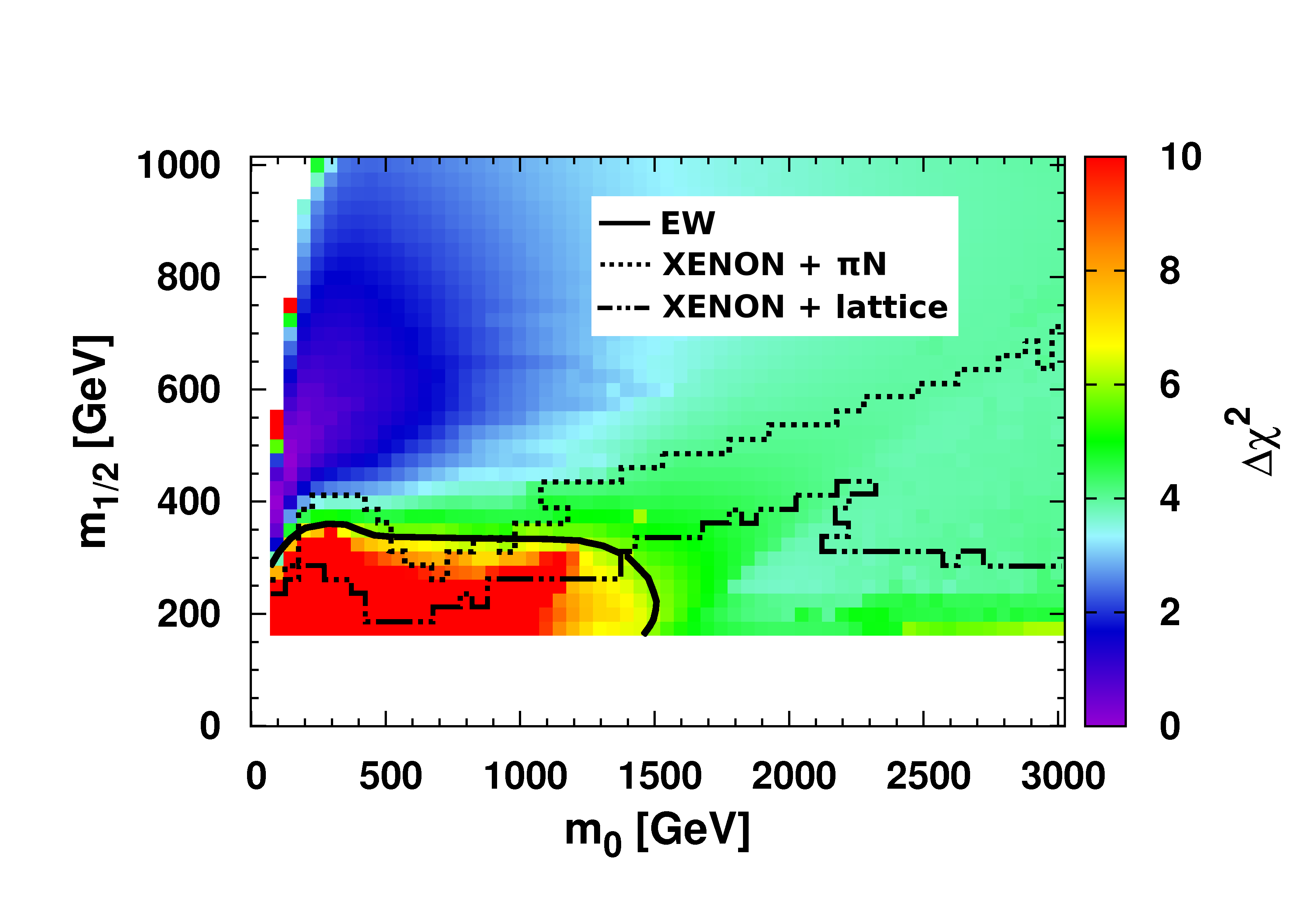}
 \includegraphics[width=0.45\textwidth]{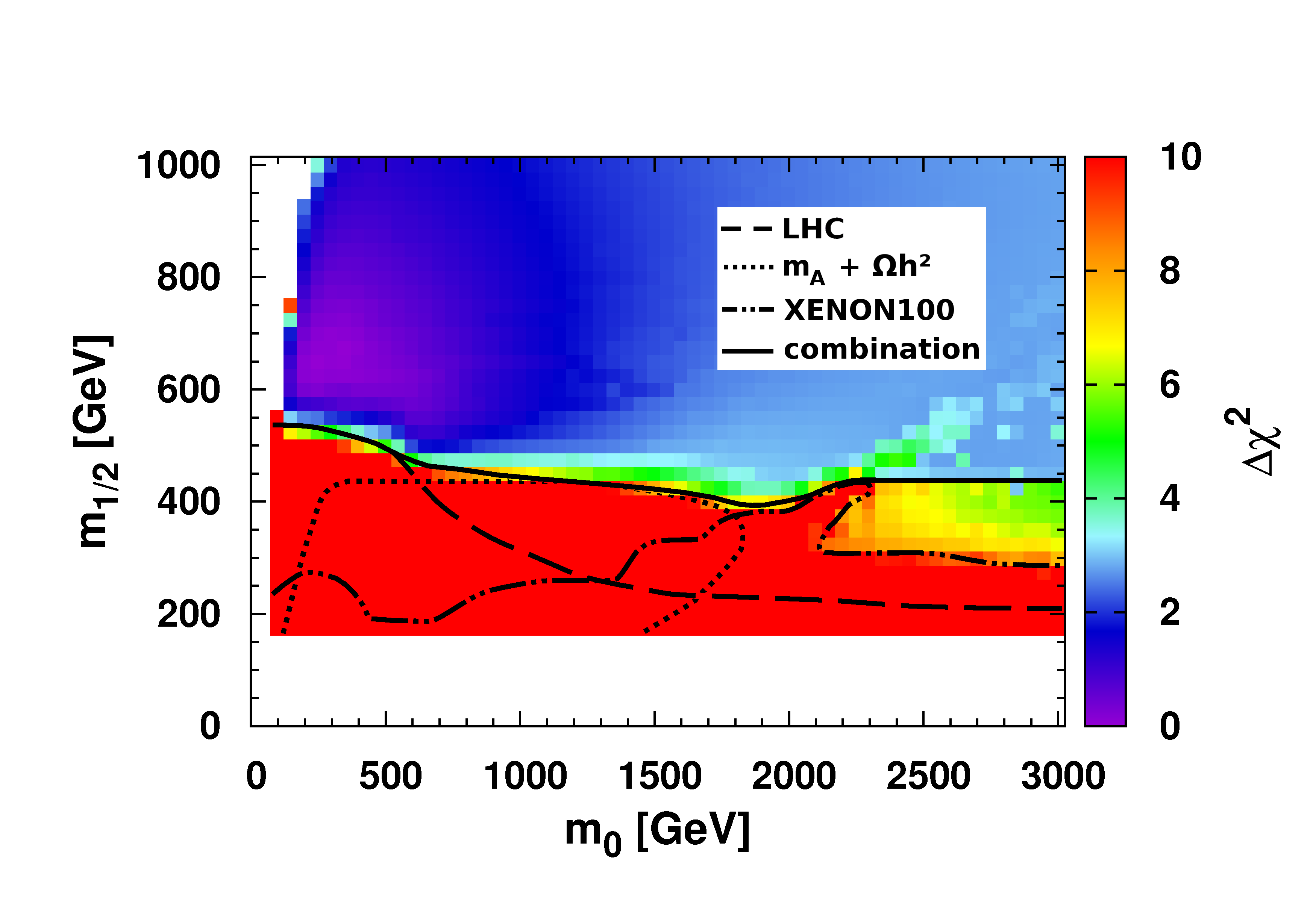}
 \end{center}
 \caption{Left: $\Delta\chi^2=\chi^2-\chi^2_{min}$ distribution in the $m_0,m_{1/2}$ plane 
 after imposing the  electroweak constraints  in comparison
with the XENON100 limits \cite{Aprile:2011hi} on the direct WIMP-nucleon cross section for two
values of the form factors (dotted line: $\pi N$ scattering, dashed dotted line:
lattice gauge theories). Values with $\Delta\chi^2=\chi^2-\chi^2_{min} >$ 5.99 are excluded (red region).
$\chi^2_{min}$=2.5 for two degrees of freedom. Right: as left after imposing the combined constraints from LHC, WMAP,
electroweak constraints and XENON100. $\chi^2_{min}$=3.6 for two degrees of freedom, which is slightly higher than in the left panel
 because of the increased contribution from g-2, see Fig. \ref{f9}. }\label{f8}
 \end{figure}

\begin{figure}
 \begin{center}
 \includegraphics[width=0.45\textwidth]{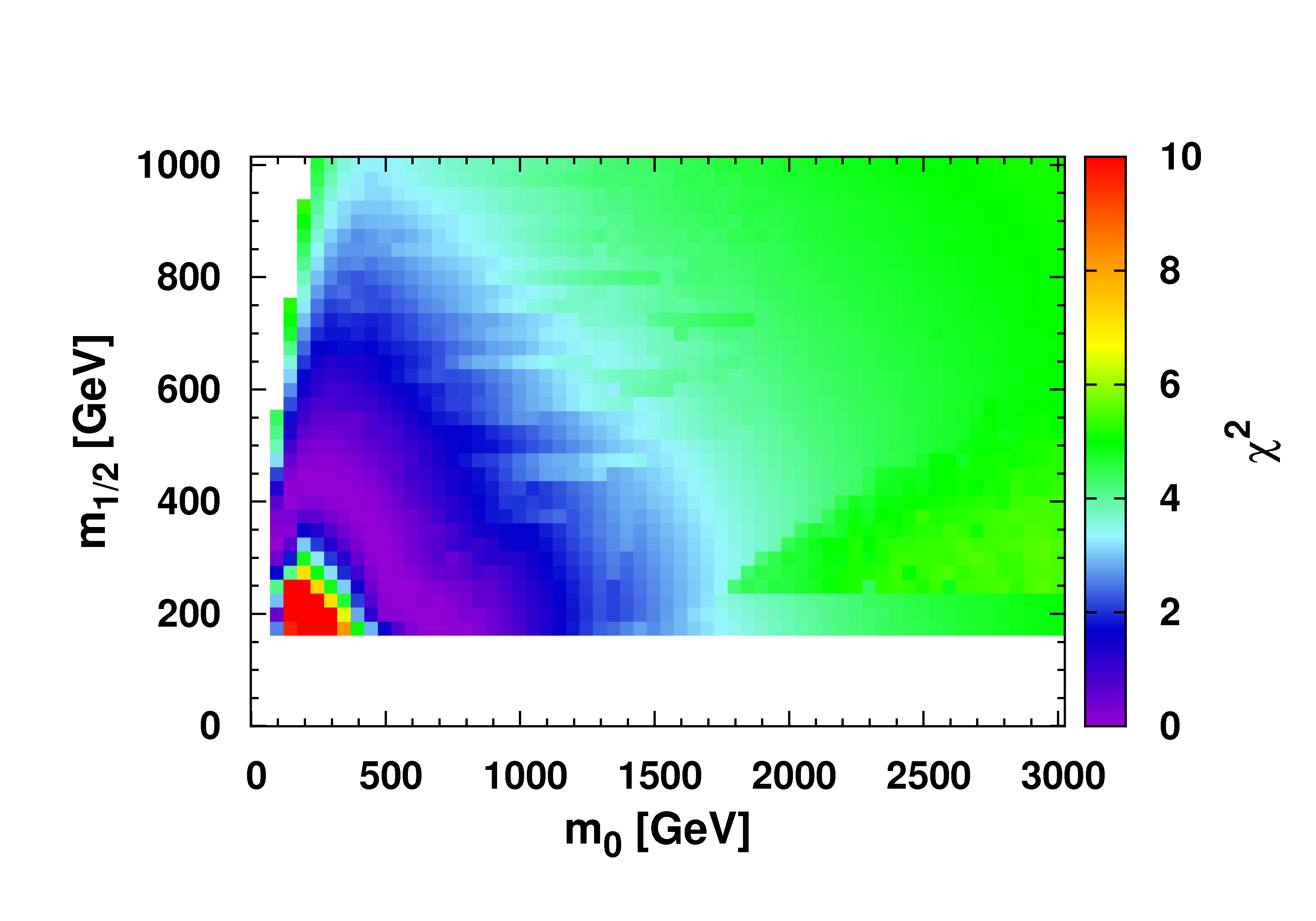}
 \includegraphics[width=0.45\textwidth]{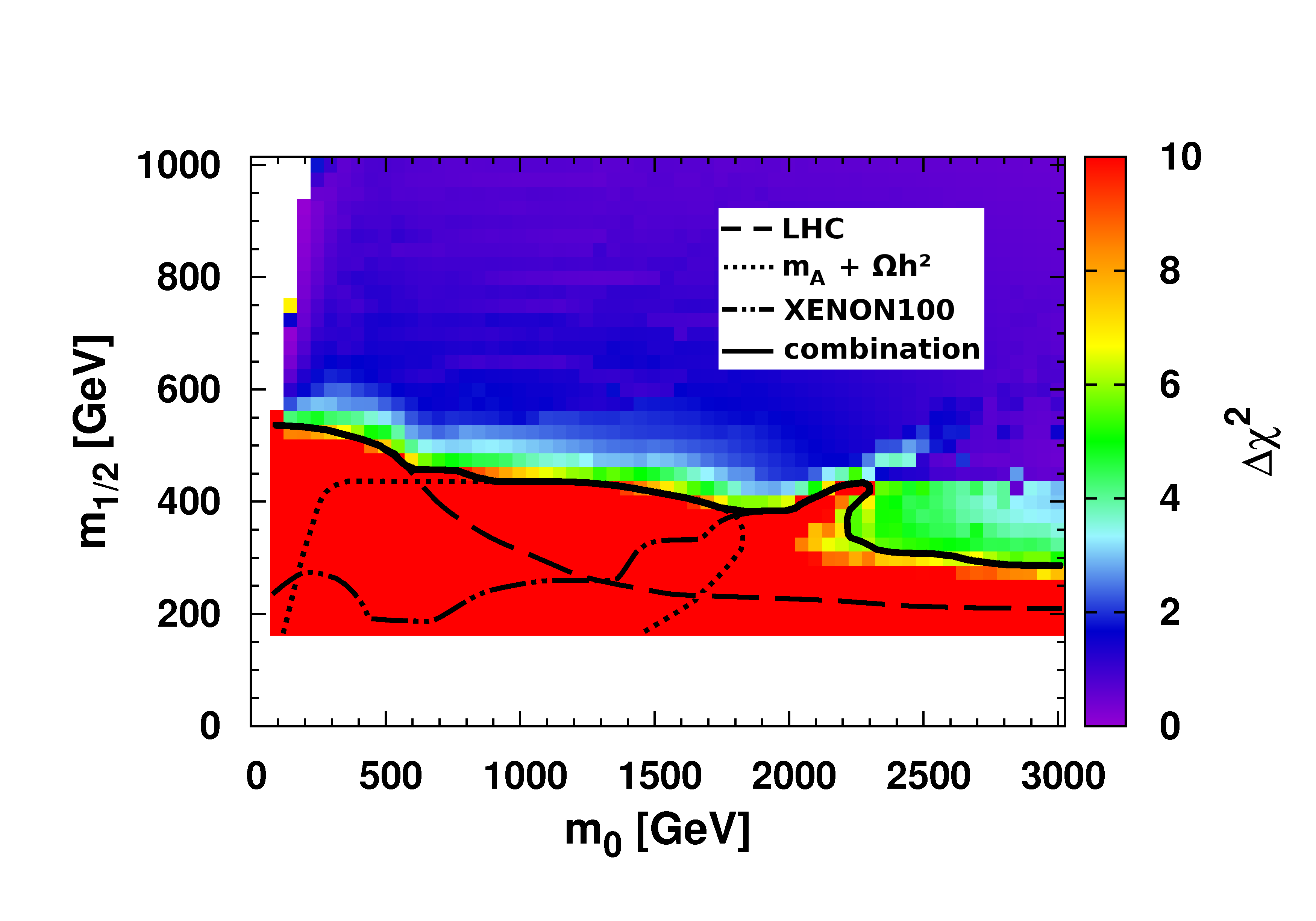}
 \end{center}
 \caption{Left: $\Delta\chi^2=\chi^2-\chi^2_{min}$ distribution of the g-2 observable alone under 
 the constraint that \tb~ and $A_0$ are still fixed by all other constraints. One observes a shallow increase of the $\chi^2$ value for large SUSY masses, because g-2 prefers light SUSY particles.
Right: the total $\Delta\chi^2$ distribution without g-2 constraint. One observes that all points above
the excluded region (solid line) are equally probable. Note that the combined limit is slightly reduced at large values of $\mzero$ in comparison with Fig. \ref{f8}, right panel, while g-2 still contributes, even if the errors are added linearly.}\label{f9}
 \end{figure}

If we exclude g-2, the remaining observables are insensitive to the  region with large SUSY masses, as can be seen from the  flat $\chi^2$ distribution in the right panel of Fig.  \ref{f9},
which can be compared with the  right panel of Fig.  \ref{f8}, where g-2 was included.
One observes that the XENON100 limit excludes $\mhalf < 300$  GeV for large $\mzero$, but if combined with  g-2  the exclusion goes up to 400 GeV in this region.

\begin{figure} 
 \begin{center}
 \includegraphics[width=0.45\textwidth]{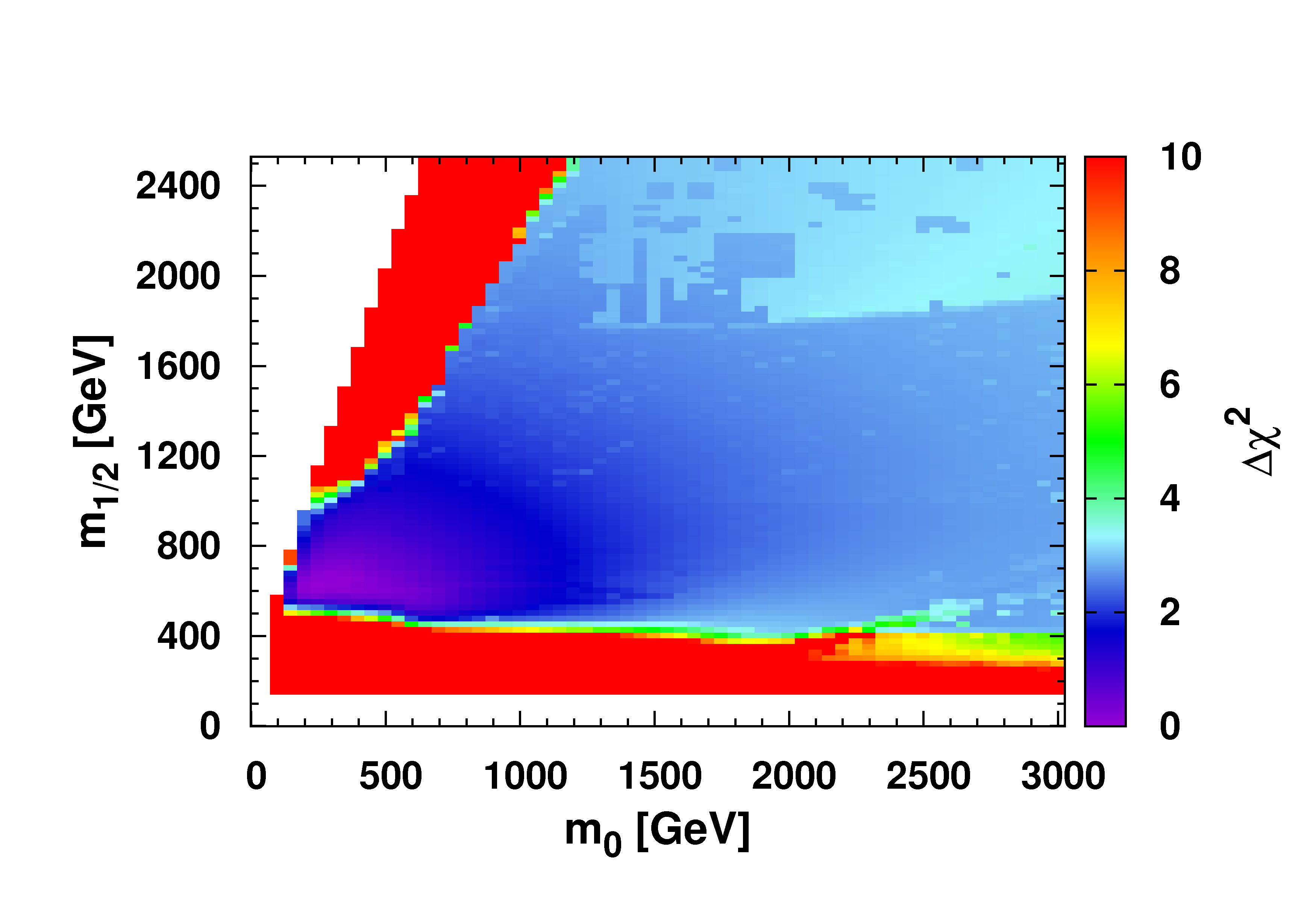}
 \includegraphics[width=0.45\textwidth]{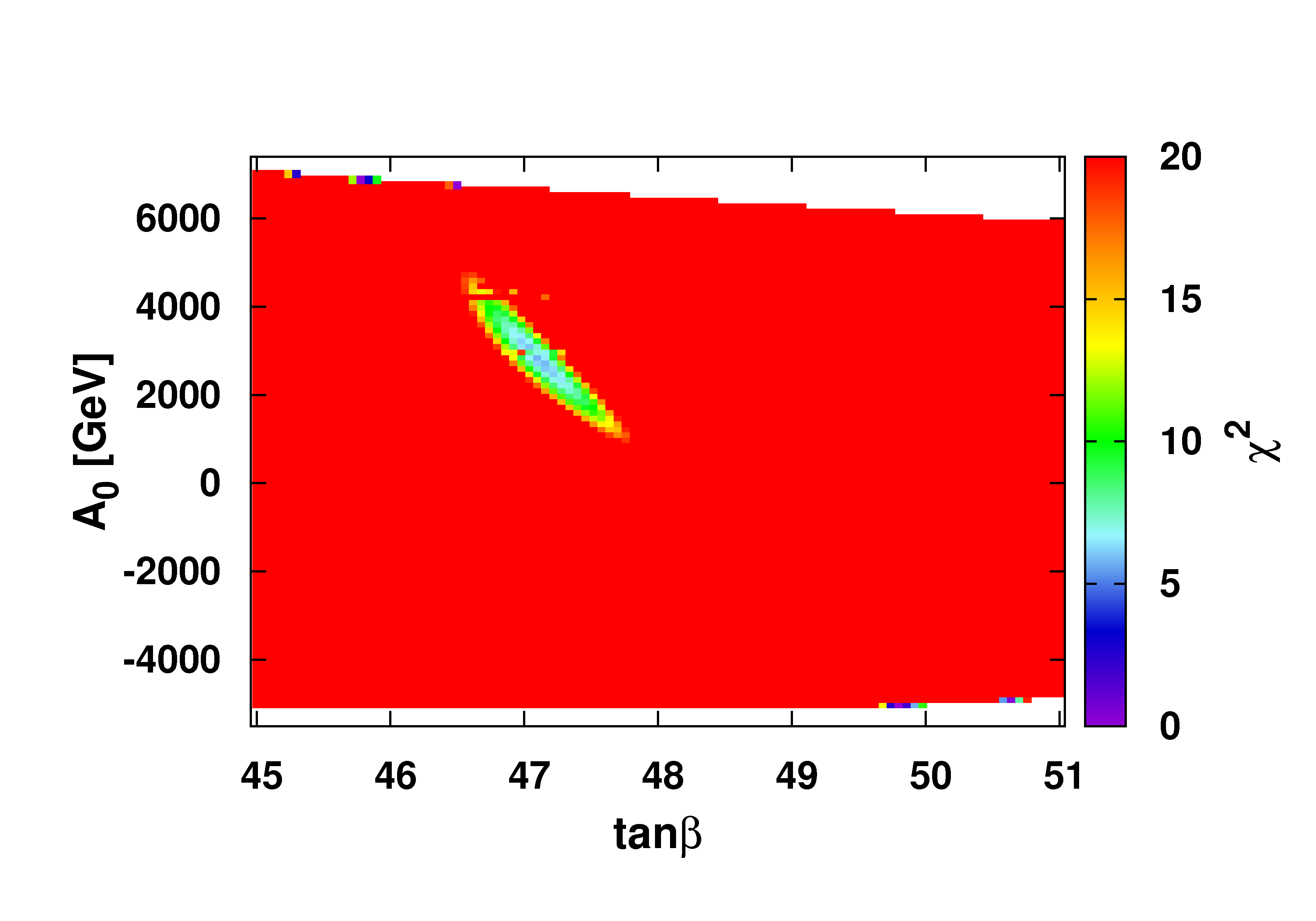}
 \end{center}
 \caption{Left: $\Delta\chi^2$ distribution of all constraints up to $\mhalf=2500$ GeV, showing that the  $\chi^2$ does not increase because of the  relic density for large neutralino masses in contrast to Ref. \cite{Buchmueller:2011sw}. The white region in the top left corner is excluded because
 the stau is always the LSP. The red region in this corner is excluded by the relic
 density constraint requiring large $\tb$, which in turn cause a large mixing in the stau sector leading
 to the  stau becoming the LSP again. Right: \tb~ and $A_0$ are strongly correlated for large neutralino masses by the relic density constraint; here $\mzero=2500$ GeV and $\mhalf=2000$ GeV has been chosen as an example. 
}\label{f10}
 \end{figure} 

\section{Summary}

If one combines the  excluded region from the direct searches at the LHC (Fig. \ref{f4}, left) and the relic density from WMAP combined with the
already stringent limits on the pseudo-scalar Higgs mass (Fig. \ref{f6}) with the XENON100 limits (Fig. \ref{f8}, left)
one obtains the excluded region shown in the right panel of Fig. \ref{f8}. Here the g-2 limit is included with the conservative linear addition of theoretical and experimental errors.
One observes that the combination  excludes $m_{1/2}$ below 400 GeV
for all values of $m_0$, which implies the LSP has to have
masses above 160 GeV within the CMSSM from the constraints considered. The
gluino mass has to be above 1 TeV in this case.

As discussed in Sect. \ref{lhc}, the LHC becomes rather insensitive to the large $\mzero$ region
because of the decreasing cross section for the production of strongly interaction particles
and the large background for the production of gauginos. However, in this region
 one obtains an increased sensitivity above the LHC limits from relic density and direct DM searches for two reasons. First,  the relic density is inversely proportional to the annihilation rate, which leads to an annihilation cross section of  the order of a few pb. Such a large cross section can only be obtained by annihilation via pseudo-scalar Higgs boson exchange in the s-channel in most of the parameter space.   The mass of the pseudo-scalar Higgs boson can be tuned everywhere in the $\mzero ,\mhalf$ plane by choosing the correct value of \tb, which has to be around 50 in this case. Such a large value of \tb~ leads to a large pseudo-scalar Higgs boson cross section at the LHC, since it is proportional to \tb$^2$. The LHC limit on the pseudo-scalar Higgs mass leads to a lower limit on $\mhalf$ of 400 GeV for intermediate values of $\mzero$, as shown in Fig. \ref{f6}.

Secondly, the cross section for direct scattering of WIMPS on nuclei has an upper limit of about 10$^{-8}$ pb, i.e. many orders of magnitude below the annihilation cross section. These cross sections are related to each other by the diagrams and kinematics. The many orders of magnitude are naturally explained in Supersymmetry by the fact that both cross sections are dominated by Higgs exchange and the fact that the Yukawa couplings to the valence quarks in the proton or neutron are negligible. Most of the scattering cross section comes from the heavier sea-quarks. However, the  density of these virtual quarks inside the nuclei is small, which is one of the reasons for the small elastic scattering cross section. In addition, the momentum transfer is small, so the propagator leads to a cross section inversely proportional to the fourth power of the Higgs mass. 
At large values of $\mzero$ EWSB forces the Higgsino component of the WIMP to increase and consequently the exchange via the  Higgs, which has an amplitude  proportional to the bino-Higgsino mixing, starts to increase. This leads to an increase in the excluded region at large $\mzero$.

As mentioned in the introduction, several groups have performed similar analysis. Our results are closest to the one of Ref. \cite{Buchmueller:2011sw}. They define the limits within the frequentist approach in a similar way as we do. Instead of optimizing the SUSY parameters by calculating the observables during execution of the program, they prepare a large data base of randomly sampled SUSY points with the observables calculated at each point. The results for small values of $\mhalf$ are similar to ours, although they use cross sections for the XENON100 results with couplings intermediate between the $\pi N$ scattering and lattice gauge theories, while we used conservatively the lattice gauge theories. The difference between $\pi N$ scattering and lattice gauge theories has been displayed in the left panel of Fig. \ref{f8}.  They display results up to $\mhalf=2500 $ GeV, since they find excluded regions above this value, which is due to the relic density constraint \cite{heinemeyer}. In our case we do find good solutions and no excluded region is found above $\mhalf=400$ GeV, as shown in Fig. \ref{f10}, left panel. This is probably due to the fact that in this region $\tb$ and $A_0$ are highly correlated, so they can be easily missed in randomly chosen SUSY samples. The strong correlation is shown in the right panel of Fig. \ref{f10} and the best solutions are obtained close to the white stripes at the top and bottom, which are near  the stau co-annihilation region. In the white region the stau is the LSP. As shown in the right panel of Fig. 10 there is no preferred region above $\mhalf=400$ GeV, if g-2 is excluded and the region where the stau becomes the LSP is ignored. The preferred minimum for g-2 (around $\mzero=400, \mhalf=200$ GeV (Fig. \ref{f9} left) is already excluded by the LHC data and the slight preference above $\mhalf=400$ is solely due to the shallow tail in the $\chi^2$ distribution of g-2 (Fig. \ref{f9}, left panel). How strong this preference is depends then on the treatment of the errors of g-2. As argued above the theoretical errors of the light-by-light scattering dominate and are certainly non-Gaussian, in which case a linear addition of the experimental and theoretical errors is the more conservative approach, so we do not think the preference by the region selected by g-2 and the corresponding preference for the expected SUSY masses is  worth emphasizing in contrast to Ref. \cite{Buchmueller:2011sw}.
  
Our results differ significantly from results using Markov Chain Monte Carlo sampling. E.g. in Ref. \cite{Bertone:2011nj} values for intermediate values of $\mzero$ are excluded, which is the region of large $\tb$ (see Fig. \ref{f6}, left panel). Here the parameters $\tb$ and $A_0$ are highly correlated again (Fig. \ref{f10}, right panel) and finding the correct minimum depends strongly on the stepping algorithm, e.g. stepping in the logarithm of a parameter is different from stepping in the parameter ("prior dependence").  
 %Ref. \cite{Sekmen:2011cz} also uses a Markov Chain, but finds again different results, especially their algorithm seems to equally sample $\tb$  for all values, which implies according to Fig. \ref{f6} that they sample most of the time the co-annihilation regions, where $\tb$ is below 50. 
Such dependence on sampling techniques largely disappears in our multistep fitting technique, since for each point of the $\mzero,\mhalf$ grid a unique solution is found independent of the minimzer used, so the frequentist approach with $\chi^2$ minimization yields the same results as a likelihood optimization with a Markov Chain sampling technique.

If one combines the limits from the direct searches at the  LHC, heavy flavour constraints, WMAP and XENON100 using the most conservative assumptions of linear addition of theoretical and experimental errors  and the lowest local relic density and matrix elements for the XENON100 limit we exclude values of $\mhalf$ below  400 GeV in the CMSSM $independent$  of $\mzero$, which implies a lower limit on the WIMP mass of 160 GeV and a gluino mass above 1 TeV. This is the CMSSM parameter region, where SUSY can be found.

 Note that the CMSSM relates electroweak gauginos and gluinos by requiring gaugino mass unification. However, the squark and gluino masses are related by the squark self-energy diagrams, which are valid in any model with couplings between squarks and gluinos. So the gluino limits from Fig. \ref{f5} of 620 GeV are model-independent.

\section{Acknowledgements} Support from the Deutsche Forschungsgemeinschaft (DFG) via a Mercator Professorship (Prof. Kazakov) and the Graduiertenkolleg  "Hochenergiephysik und Teilchenastrophysik" in Karlsruhe  is greatly appreciated. Furthermore, support from the Bundesministerium for Bildung und Forschung (BMBF) is acknowledged.
We thank A. Phukov for help in calculating the SUSY cross section with CalcHEP inside   micrOMEGAs.

%\bibliographystyle{lucas_unsrt}
%\bibliography{mybib}
%\end{document}

\input{xenon100a.bbl}
\end{document}

%% file: config.tex
\newlength{\dslashwidth}

\newcommand{\bsg}{\ensuremath{b\to X_s\gamma}}

\newcommand{\tb}{\ensuremath{\tan\beta}}

\newcommand{\beq}{\begin{equation}} 
\newcommand{\eeq}{\end{equation}}
\newcommand{\beqa}{\begin{eqnarray}} 
\newcommand{\eeqa}{\end{eqnarray}}
\newcommand{\newc}{\newcommand}
\newcommand{\bq}{\begin{equation}}
\newcommand{\eq}{\end{equation}}
\newcommand{\ba}{\begin{array}}
\newcommand{\ea}{\end{array}}
\newcommand{\bqa}{\begin{eqnarray}}
\newcommand{\eqa}{\end{eqnarray}}

\newcommand{\lnf}{{\ifmmode \Lambda^{(N_f)} \else $\Lambda^{(N_f)}$\fi}}
\newcommand{\ms}{{\ifmmode \overline{MS} \else $\overline{MS}$\fi}}
\newcommand{\dr}{{\ifmmode \overline{DR} \else $\overline{DR}$\fi}}
\newcommand{\lms}{{\ifmmode \Lambda^{(5)}_{\overline{MS}} \else $\Lambda^{(5)}_{\overline{MS}}$\fi}}
\newcommand{\lam}{{\ifmmode \Lambda \else $\Lambda$\fi}}
\newcommand{\mev}{{\ifmmode {\rm MeV} \else ${\rm MeV}$\fi}}
\newcommand{\gev}{{\ifmmode {\rm GeV} \else ${\rm GeV}$\fi}}
\newcommand{\gevc}{{\ifmmode {\rm GeV/c^2} \else ${\rm GeV/c^2}$\fi}}
\newcommand{\tev}{{\ifmmode {\rm TeV} \else ${\rm TeV}$\fi}}
\newcommand{\tevc}{{\ifmmode {\rm TeV/c^2} \else ${\rm TeV/c^2}$\fi}}
\newcommand{\lp}{{\ifmmode L^+  \else $L^+$\fi}}
\newcommand{\lm}{{\ifmmode L^-  \else $L^-$\fi}}
\newcommand{\mlp}{{\ifmmode M(L^-) \else $M(L^-)$\fi}}
\newcommand{\mlz}{{\ifmmode M(L^0) \else $M(L^0)$\fi}}
\newcommand{\lz}{{\ifmmode L^0 \else $L^0$\fi}}
\newcommand{\ev}{{\ifmmode GeV/c^2 \else $GeV/c^2$\fi}}
\newcommand{\tri}{{\ifmmode \triangleup \else $\triangleup$\fi}}
\newcommand{\unl}{{\ifmmode U_{lL^0} \else $U_{lL^0}$\fi}}\newcommand{\gL}{{\ifmmode g_L \else $g_{L}$\fi}}
\newcommand{\gR}{{\ifmmode g_R  \else $g_{R}$\fi}}
\newcommand{\gumu}{{\ifmmode \gamma^{\mu} \else $\gamma^{\mu}$\fi}}
\newcommand{\gunu}{{\ifmmode \gamma^{\nu} \else $\gamma^{\nu}$\fi}}
\newcommand{\gdmu}{{\ifmmode \gamma_{\mu} \else $\gamma_{\mu}$\fi}}
\newcommand{\gdnu}{{\ifmmode \gamma_{\nu} \else $\gamma_{\nu}$\fi}}
\newcommand{\stw}{{\ifmmode\sin^2\theta_W \else $\sin^{2}\theta_{W}$ \fi}}
\newcommand{\sws}{{\ifmmode \;\sin^2\theta_W  \else $\;\sin^{2}\theta_{W}$ \fi}}
\newcommand{\cws}{{\ifmmode \;\cos^2\theta_W  \else $\;\cos^{2}\theta_{W}$ \fi}}
\newcommand{\sw}{{\ifmmode \;\sin\theta_W  \else $\sin\theta_{W}$ \fi}}
\newcommand{\cw}{{\ifmmode \;\cos\theta_W  \else $\;\cos\theta_{W}$ \fi}}
\newcommand{\tw}{{\ifmmode \;\tan\theta_W  \else $\;\tan\theta_{W}$ \fi}}
\newcommand{\qq}{{\ifmmode q\overline{q} \else $q\overline{q}$\fi}}
\newcommand{\lR}{{\ifmmode l_R \else $l_R$\fi}}
\newcommand{\lL}{{\ifmmode l_L \else $l_L$\fi}}
\newcommand{\nt}{{\ifmmode \nu_{\tau} \else $\nu_{\tau}$\fi}}
\newcommand{\nuR}{{\ifmmode \nu_R  \else $\nu_R$\fi}}
\newcommand{\nuL}{{\ifmmode \nu_L  \else $\nu_L$\fi}}
\newcommand{\qR}{{\ifmmode g_R  \else $q_R$\fi}}
\newcommand{\qL}{{\ifmmode q_L  \else $q_L$\fi}}
\newcommand{\qRp}{{\ifmmode q_R'  \else $q_{R}$'\fi}}
\newcommand{\qLp}{{\ifmmode q_L'  \else $q_{L}$'\fi}}
\newcommand{\est}{{\ifmmode e^{\bf \ast} \else $e^{\bf \ast}$\fi}}
\newcommand{\lst}{{\ifmmode l^{\bf \ast} \else $l^{\bf \ast}$\fi}}
\newcommand{\must}{{\ifmmode \mu^{\bf \ast} \else $\mu^{\bf \ast}$\fi}}
\newcommand{\taust}{{\ifmmode \tau^{\bf \ast} \else $\tau^{\bf \ast}$ \fi}}
\newcommand{\pperp}{{\ifmmode p_t  \else $p_t$\fi}}
\newcommand{\et}{{\ifmmode E_t  \else $E_t$\fi}}
\newcommand{\xt}{{\ifmmode x_t  \else $x_t$\fi}}
\newcommand{\smumu}{{\ifmmode \sigma_{\mu\mu}  \else $\sigma_{\mu\mu}$ \fi}}
\newcommand{\eg}{{\ifmmode e\gamma  \else $e\gamma$\fi}}
\newcommand{\epem}{{\ifmmode e^+e^-  \else $e^+e^-$\fi}}
\newcommand{\lplm}{{\ifmmode L^+L^-  \else $L^+L^-$\fi}}
\newcommand{\pp}{{\ifmmode p\overline p  \else $p\overline p$\fi}}
\newcommand{\llz}{{\ifmmode L^0\overline{L}^0 \else $L^0\overline{L}^0$\fi}}
\newcommand{\epemt}{{\ifmmode e^+e^- \to  \else $e^+e^- \to$\fi}}
\newcommand{\eb}{{\ifmmode E_{beam}  \else $E_{beam}$\fi}}
\newcommand{\ip}{{\ifmmode pb^{-1}  \else $pb^{-1}$\fi}}
\newcommand{\upm}{{\ifmmode ^{\pm}  \else $^{\pm}$\fi}}
\newcommand{\de}{{\ifmmode ^{\circ}  \else $^{\circ}$ \fi}}
\newcommand{\appr}{{\ifmmode \sim \else $\sim$ \fi}}
\newcommand{\corresp}{{\ifmmode \stackrel{\wedge}{=} \else $\stackrel{\wedge}{=}$ \fi}}
\newcommand{\sqrts}{{\ifmmode \sqrt{s} \else $\sqrt{s}$\fi}}
\newcommand{\zz}{{\ifmmode Z^0  \else $Z^0$\fi}}
\newcommand{\mz}{{\ifmmode M_{Z}  \else $M_{Z}$\fi}}
\newcommand{\mzs}{{\ifmmode M_{Z}^2  \else $M_{Z}^2$\fi}}
\newcommand{\mw}{{\ifmmode M_{W}  \else $M_{W}$\fi}}
\newcommand{\mws}{{\ifmmode M_{W}^2  \else $M_{W}^2$\fi}}
\newcommand{\mh}{{\ifmmode M_{Higgs}  \else $M_{Higgs}$\fi}}
\newcommand{\gt}{{\ifmmode \Gamma_{tot} \else $\Gamma_{tot}$\fi}}
\newcommand{\msusy}{{\ifmmode M_{SUSY}  \else $M_{SUSY}$\fi}}
\newcommand{\msusys}{{\ifmmode M_{SUSY}^2  \else $M_{SUSY}^2$\fi}}
\newcommand{\su}{{\ifmmode SU(3)_C\otimes\- SU(2)_L\otimes\- U(1)_Y \else $SU(3)_C\otimes SU(2)_L\otimes U(1)_Y$\fi}}
\newcommand{\suthree}{{\ifmmode SU(3)_C  \else $SU(3)_C$\fi}}
\newcommand{\sutwo}{{\ifmmode  SU(2)_L\otimes U(1)_Y \else $SU(2)_L\otimes U(1)_Y$\fi}}
\newcommand{\taup}{{\ifmmode \tau_{proton} \else $\tau_{proton}$\fi}}
\newcommand{\as}{{\ifmmode \alpha_{s}  \else $\alpha_{s}$\fi}}
\newcommand{\mgut}{{\ifmmode M_{GUT}  \else $M_{GUT}$\fi}}
\newcommand{\mguts}{{\ifmmode M_{GUT}^2  \else $M_{GUT}^2$\fi}}
\newcommand{\mzero}{{\ifmmode m_0        \else $m_0$\fi}}
\newcommand{\mhalf}{{\ifmmode m_{1/2}    \else $m_{1/2}$\fi}}
\newcommand{\sq}{{\ifmmode \tilde{q}    \else $\tilde{q}$\fi}}
\newcommand{\gl}{{\ifmmode \tilde{g}    \else $\tilde{g}$\fi}}
\newcommand{\mb}{{\ifmmode m_{b}    \else $m_{b}$\fi}}
\newcommand{\mt}{{\ifmmode m_{t}    \else $m_{t}$\fi}}
\newcommand{\mts}{{\ifmmode m_{t}^2    \else $m_{t}^2$\fi}}

\newcommand{\mtau}{{\ifmmode m_{\tau}  \else $m_{\tau}$\fi}}
\newcommand{\dpp}{{\ifmmode \delta_{pert} \else $\delta_{pert}$\fi}}
\newcommand{\dnp}{{\ifmmode\delta_{non-pert}\else$\delta_{non-pert}$\fi}}
\newcommand{\dew}{{\ifmmode \delta_{\rm EW}\else $\delta_{\rm EW}$\fi}}
\newcommand{\rt}{{\ifmmode R_{\tau}  \else $R_{\tau} $\fi}}
\newcommand{\rz}{{\ifmmode R_{Z}  \else $R_{Z} $\fi}}

\newcommand{\swb}{{\ifmmode \sin^2\theta_{\overline{MS}} \else $\sin^2\theta_{\overline{MS}}$\fi}}
\newcommand{\cwb}{{\ifmmode \cos^2\theta_{\overline{MS}} \else $\cos^2\theta_{\overline{MS}}$\fi}}

\newcommand{\bsmm}{\ensuremath{B^0_s\to\mu^+\mu^-}}

\newcommand{\btaunu}{\ensuremath{B_u\to\tau\nu}}

\newc\AIPCP[3] {{\em AIP Conf. Proc.} {\bf #1} (#2) #3}
\newc\AJ[3] {{\em Astrophys. J.} {\bf #1} (#2) #3}
\newc\AMS[3] {{\em Ann. Math. Statist.} {\bf #1} (#2) #3}
\newc\AP[3] {{\em Ann. Phys.} {\bf #1} (#2) #3}
\newc\APJ[3] {{\em Astropart. J.} {\bf #1} (#2) #3}
\newc\APP[3] {{\em Astropart. Phys.} {\bf #1} (#2) #3}
\newc\APS[3] {{\em Astrophys. J. Suppl.} {\bf #1} (#2) #3}
\newc\ARNPS[3] {{\em Ann. Rev. Nucl. Part. Sci.} {\bf C#1} (#2) #3}
\newc\BA[3] {{\em Bayesian Anal.} {\bf C#1} (#2) #3}
\newc\CPC[3] {{\em Comput. Phys. Commun.} {\bf C#1} (#2) #3}
\newc\CP[3] {{\em Contemp. Phys.} {\bf #1} (#2) #3}
\newc\EPJ[3] {{\em Euro. Phys. Journ.} {\bf C#1} (#2) #3}
\newc\JCAP[3] {{\em JCAP} {\bf #1} (#2) #3}
\newc\JHEP[3] {{\em JHEP} {\bf #1} (#2) #3}
\newc\JPG[3] {{\em J. Phys.} {\bf G #1} (#2) #3}
\newc\IJMP[3] {{\em Int. J. Mod. Phys.} {\bf A #1} (#2) #3}
\newc\MNRAS[3] {{\em Mon. Not. Roy. Astron. Soc.} {\bf #1} (#2) #3}
\newc\MPL[3] {{\em Mod. Phys. Lett.} {\bf A #1} (#2) #3}
\newc\NAR[3] {{\em New Astron. Rev.} {\bf #1} (#2) #3}
\newc\NCA[3] {{\em Nuovo Cimento} {\bf #1} (#2) #3}
\newc\NIM[3] {{\em Nucl. Instrum. Methods} {\bf #1} (#2) #3}
\newc\NIMA[3] {{\em Nucl. Instrum. Methods} {\bf A #1} (#2) #3}
\newc\NAT[3] {{\em Nature} {\bf #1} (#2) #3}
\newc\NPB[3] {{\em Nucl. Phys.} {\bf B #1} (#2) #3}
\newc\NPA[3] {{\em Nucl. Phys.} {\bf A #1} (#2) #3}
\newc\NPPS[3] {{\em Nucl. Phys. Proc. Suppl.} {\bf #1} (#2) #3}
\newc\PLB[3] {{\em Phys. Lett.} {\bf B #1} (#2) #3}
\newc\PR[3] {{\em Phys. Rep.} {\bf #1} (#2) #3}
\newc\PRL[3] {{\em Phys. Rev. Lett.} {\bf #1} (#2) #3}
\newc\PRD[3] {{\em Phys. Rev.} {\bf D #1} (#2) #3}
\newc\PRC[3] {{\em Phys. Rev.} {\bf C #1} (#2) #3}
\newc\PTP[3] {{\em Prog. Theor. Phys.} {\bf #1} (#2) #3}
\newc\RMP[3] {{\em Rev. Mod. Phys.} {\bf #1} (#2) #3 }
\newc\RPP[3] {{\em Rept. Prog. Phys.} {\bf #1} (#2) #3 }
\newc\SC[3] {{\em Science} {\bf #1} (#2) #3 }
\newc\ZPC[3] {{\em Z. Phys.} {\bf C #1} (#2) #3}
\newc\Err[3] {{\em Erratum-ibid.} {\bf #1} (#2) #3 }

%% file: xenon100a.bbl
\providecommand{\href}[2]{#2}\begingroup\raggedright\endgroup